\documentclass[%
twocolumn,
 amsmath,amssymb,
prb,
floatfix,
]{revtex4-2}
\pdfoutput=1
\usepackage{hyperref}
\usepackage{xcolor}

\usepackage{subfigure}
\usepackage[normalem]{ulem}
\usepackage{graphicx, }
\usepackage{dcolumn}
\usepackage{bm}

\begin{document}


    \title{Multiscale modelling of magnetostatic effects on magnetic nanoparticles with application to hyperthermia
    }





\author{Razyeh Behbahani$^{1, 2}$, Martin L.~Plumer$^1$ and Ivan Saika-Voivod$^{1, 2}$}
\address{$^1$ Department of Physics and Physical Oceanography, Memorial University of Newfoundland, St. John's, Newfoundland and Labrador, Canada, A1B 3X7}

 \address{$^2$ Department of Applied Mathematics, University of Western Ontario, London, Ontario, Canada, N6A 3K7
}

\date{\today}

\begin{abstract}

We extend a renormalization group-based course-graining method for micromagnetic simulations to include properly scaled magnetostatic interactions.  We apply the method in simulations of dynamic hysteresis loops at clinically relevant sweep rates and at 310 K of iron oxide nanoparticles (NPs) of the kind that have been used in preclinical studies of magnetic hyperthermia. The coarse-graining method, along with a time scaling involving sweep rate and Gilbert damping parameter, allow us to span length scales from the unit cell to NPs approximately 50~nm in diameter with reasonable simulation times.  For both NPs and the nanorods composing them, we report effective uniaxial anisotropy strengths and saturation magnetizations, which differ from those of the bulk materials magnetite and maghemite of which they are made, on account of the combined non-trivial effects of temperature, inter-rod exchange, magnetostatic interactions and the degree of orientational order within the nanorod composites. The effective parameters allow treating the NPs as single macrospins, and we find for the test case of calculating loops for two aligned NPs that using the dipole approximation is sufficient for distances beyond 1.5 times the NP diameter.  We also present a study on relating integration time step to micromagnetic cell size,  finding that the optimal time step size scales approximately linearly with cell volume.





\end{abstract}

\maketitle

\section{\label{sec:intro}Introduction}
    

The use of micromagnetics based on the Landau-Lifshitz-Gilbert (LLG) equations for the simulation of dynamic hysteretic magnetization-magnetic field (MH) loops at room temperature and at kHz frequencies relevant for magnetic hyperthermia applications offers a challenging  area of the study for coarse graining. For numerical studies based on micromagnetics, hysteretic heating is typically associated with the specific loss power (SLP) and is assumed to be proportional to the area of a calculated MH loop. In a recent work~\cite{BehCoarse-graining2020} (hereafter referred to as I), we employed and modified a renormalization group (RG) approach introduced by Grinstein and Koch~\cite{grinstein2003coarse} for our model system of magnetite (Fe$_3$O$_4$) nanorods that form the building blocks of nanoparticles used in preclinical magnetic hyperthermia trials on mice \cite{dennis2009nearly}.  Our study focused on MH loops and demonstrated that for the case of individual nanorods, where exchange interactions, uniaxial anisotropy, and a sinusoidal external field are included in the model of uniformly magnetized cells, the RG approach works well over an entire range of fixed-volume rods composed of from 10752 cells ($b=1$) to one cell ($b=22$), where the smallest cell size of the scaling parameter $b=1$ corresponds to the dimensions of the magnetite unit cell.  Our work also illustrates that significant additional computational speed-up can be achieved over the dynamic range of interest by maintaining a constant value for SR/$\alpha$, where SR is the designated sweep rate (in units of Oe/s) of the MH loop simulation and $\alpha$ is the LLG damping constant.  This work, which employed OOMMF micromagnetics software~\cite{OOMMF}, omitted explicit magnetostatic interactions but these were accounted for through an effective uniaxial anisotropy. 

Here, our previous work is extended with several objectives.  The first is to develop a coarse-graining algorithm for dynamic MH loops for a single nanorod that has explicit magnetostatic interactions included (in addition to the scaling of the magnetization, exchange, anisotropy and applied field used previously), which were omitted in the RG analysis of Grinstein and Koch~\cite{grinstein2003coarse}. This study allows for the estimation of an effective single-ion anisotropy that mimics the effects of the self-demag field.  The second goal is to examine MH loops corresponding to magnetic nanoparticles (NPs) that are constructed from the nanorods where the impact of inter-rod exchange and inter-rod magnetostatic interactions are examined. This part of the study examines the case of just two adjacent nanorods in various geometries, and finishes with composites of 10 stacked rods, 
inspired by the experimental study of Dennis~et al.~\cite{dennis2009nearly}.  Different stackings represent varying degrees of orientational order of nanorods within a NP.  Loops corresponding to a variety of applied field orientations are examined.  
The third goal is to find the effective magnetization and anisotropy that allows the modelling of a NP as a single macrospin, both in the case of a single NP in a field and for two interacting NPs.  This macrospin approximation may be useful for further study of NP assemblies.
In addition, the impact of cell size on the assigned time step in the OOMMF LLG  solver is studied, where a larger time step can be used with larger cell sizes resulting in an additional increase in computational efficiency.

Magnetic hyperthermia as a novel and developing cancer treatment method continues to attract considerable attention at the applied as well as fundamental level ~\cite{shi2019enhanced, pearce2013magnetic, munoz2017towards, mehdaoui2013increase, dennis2013physics, allia2019nonharmonic}. A wide range of preclinical studies have been reported using magnetic hyperthermia as a primary or secondary cancer treatment along with conventional chemotherapy or radiotherapy~\cite{chang2018biologically, dennis2009nearly, thiesen2008clinical, sadhukha2013inhalable}. Moreover, recent analytical and numerical studies~\cite{usov2012dynamics, simeonidis2016situ, anand2016spin, torche2020thermodynamics, serantes2014multiplying, mehdaoui2013increase, anandhi2020factors} reflect the growing need for understanding the heating mechanisms of magnetic hyperthermia to provide a more accurate guide for experiments.

In magnetic hyperthermia, injected magnetic nanoparticles exhibit hysteresis under applied magnetic field and heat up and damage cancerous tumor cells. As nanoparticles are mobile inside the tumor upon injection, exploring the effects of interactions between magnetic particles, as well as possible heating mechanisms such as Brownian rotation or hysteresis heating (N\'eel relaxation), is crucial for understanding particle clustering and heating efficiency. To this end,
many studies have investigated the impact of long range dipolar interactions on hyperthermia with interesting and related results ~\cite{anand2016spin, landi2014role, haase2012role, cabrera2018dynamical,  serantes2014multiplying, mehdaoui2013increase, wu2017magnetic}.
For example, Anand~et al.~\cite{anand2016spin} examined the effect of dipole interaction strength on the heating efficiency of micron-sized particles and showed that there is an optimal NP volume fraction for maximizing SLP. Haase and Nowak~\cite{haase2012role} reported a negative effect of dipolar interactions on SLP at high particle concentrations. By contrast, Landi~\cite{landi2014role} used a mean field theory and found that the dipole interactions increase the energy barrier between stable configurations of the magnetization. He deduced that dipolar interactions improve SLP as long as certain conditions of the energy barrier of the system are met.
Such studies motivate a bottom-up approach to determining and modeling effective interparticle interactions, and underline the importance of including magnetostatic interactions in our scaling approach.


This paper is organized as follow. Our model is described in section II. Section III summarizes the coarse-graining scheme we use and in section IV we test the scaling method for multiple nanorods. Section V contains a more detailed investigation of the effect of inter-rod exchange and magnetostatic interactions to examine their effect on magnetization dynamics in a system of two nanorods. In section VI, three nanorod composites of varying internal orientational order are introduced and their effective macrospin parameters are determined. In section VII we study the hysteresis loops of 2 NP as a function of separation, and test the macrospin models in this context.  Finally, we present our conclusions in Section VIII.  As choosing the proper time step for simulating a system of study is another challenging detail in such numerical studies~\cite{lopez2012micromagnetic, kapoor2006effect}, we address it for our system in Appendix A. 
\section{\label{sec:model}The model}
\begin{figure}
    \includegraphics[width=\columnwidth ]{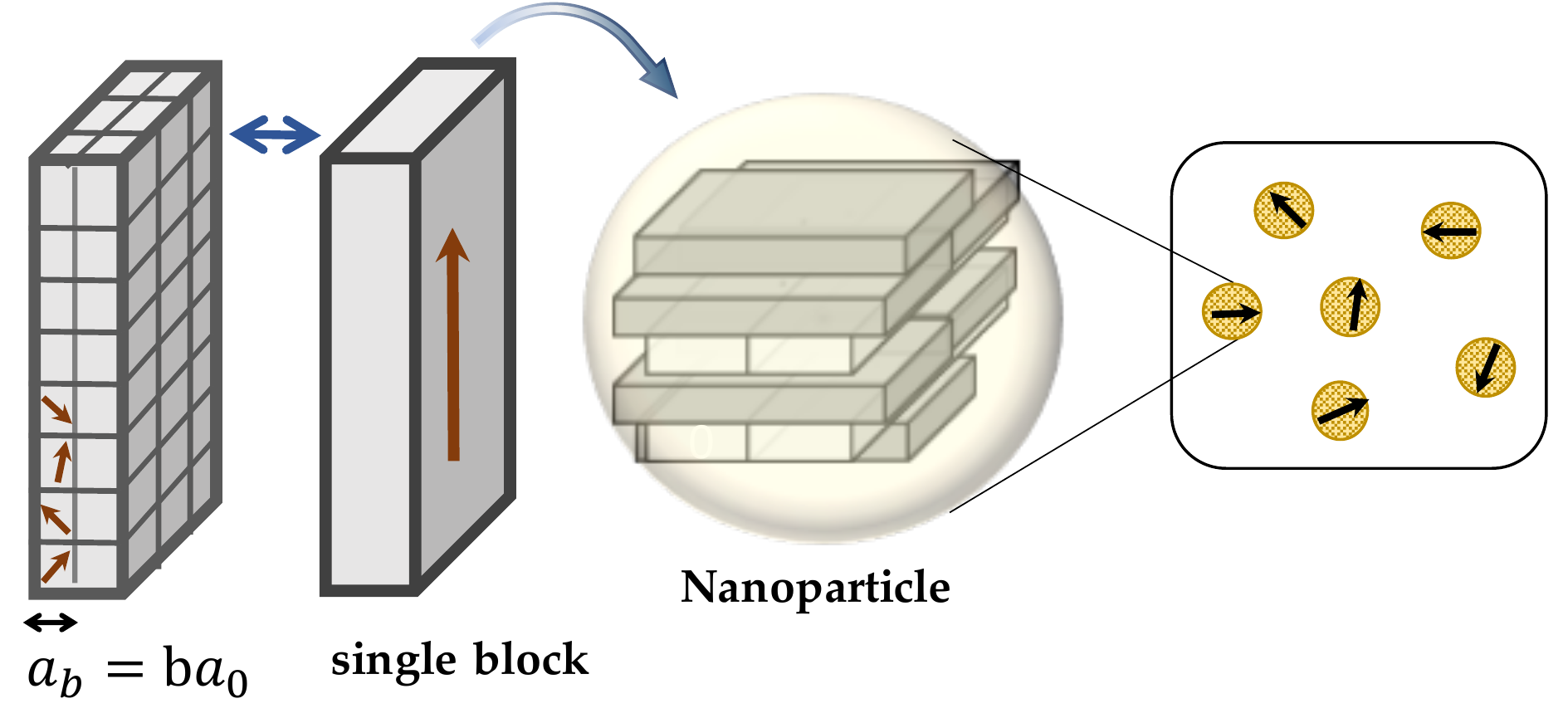}
    \caption{Coarse-graining model of a magnetite nanorod.  The smallest micromagnetic cell corresponds to the cubic unit cell of length $a_0=0.839$~nm with ferrimagnetic atomic spins represented by single magnetic moment. Larger cells are characterized by a length $a_b=b\,a_0$ for $b>1$. The number of cells is reduced from $56\times24\times8=10752$ to $N_b$=10752/$b^3$ =1344, 168 and 21 for $b=$2, 4 and 8 respectively.  A single block corresponds to $b=22$. Nanoparticles are made of nanorods.}\label{fig:micromagnetics}
\end{figure}
We wish to simulate iron oxide nanorods made of magnetite or maghemite ($\gamma$-Fe$_2$O$_3$), while including magnetostatic interactions. These two iron oxides have similar magnetic parameters, with the exception of crystalline anisotropy, which is cubic in magnetite and uniaxial for maghemite. Our research is inspired by experimental results reported by Dennis et al.~\cite{dennis2009nearly}, in which simulated nanorods are the building blocks of nanoparticles (see Fig.~2 therein).  We study here assemblies of nanorods from two to bundles of ten as single nanoparticles to explore their collective heating behaviour by calculating hysteresis loops.


For simulating nanorods with nominal dimension 6.7 nm $\times$ 20 nm $\times$ 47 nm (Fig.~\ref{fig:micromagnetics}), we use the Object Oriented MicroMagnetic Framework (OOMMF)~\cite{OOMMF}, and the smallest simulation cell we use has the dimensions of the unit cell of ferrimagnetic magnetite, represented by a single magnetization vector. We employ the Theta Evolve module~\cite{theta_evolve} required for simulations at finite $T$. The Landau-Lifshitz-Gilbert (LLG) equation is commonly used to describe the dynamics of magnetic moments~\cite{cullity2011introduction, gilbert2004phenomenological, brown1963thermal} by describing the precession and damping of a cell's magnetic moment in an effective field. The value of damping constant $\alpha$, representative of energy dissipation, for magnetite films has been reported in a range from 0.03 to 0.2 depending on the thickness~\cite{serrano2011thickness}.  Setting $\alpha$=0.1 for our system size is consistent with other reported micromagnetic studies~\cite{plumer2010micromagnetic, usov2010low}. The effective field combines Zeeman, exchange, magnetocrystalline anisotropy and magnetostatic terms.  
Additionally, Brown~\cite{brown1963thermal} provided a formalism to add thermal effects into the calculations via a random effective field. It is known that thermal fluctuations are more pronounced for smaller simulation volumes prone to superparamagnetism and simulation results strongly depend on cell size~\cite{grinstein2003coarse, lopez2012micromagnetic, lee2004excitations}.  We explore the cell size and time step correlation in Appendix~\ref{app:time_stp} for simulations at finite $T$.

As in I, we use the bulk magnetite parameters with a saturation magnetization $M_s=480$~kA/m~\cite{dutz2013magnetic, usov2013properties, heider1988note} and  exchange stiffness constant $A_0=0.98 \times 10^{-11}$~J/m~\cite{heider1988note, kouvel1956specific, moskowitz1987theoretical, glasser1963spin, srivastava1979exchange, srivastava1987spin, uhl1995first} which leads to the critical temperature of $T_c=858$~K for its cubic unit cell size $a_0=0.839$~nm. Magnetite (Fe$_3$O$_4$) possesses cubic crystalline anisotropy~\cite{shi2019enhanced, plumer2010micromagnetic, usov2013properties, abe1976magnetocrystalline, vreznivcek2012magnetocrystalline}, and as it has only a weak tendency to produce hysteresis, we omit it in magnetite simulations. However, nanorods may contain significant amounts of maghemite with uniaxial crystalline anisotropy with energy density of $K_0$=10 kJ/m$^3$ ~\cite{shi2019enhanced, plumer2010micromagnetic, shokrollahi2017review}, used in maghemite simulations in the present study.  Otherwise, we use the same parameters for maghemite as for magnetite. 

To restrict the uncontrolled heat generated by Eddy currents, the product of amplitude $\times$ frequency of the AC magnetic field should be less than a threshold that limits the sweep rate of the applied AC field to SR=4$H_{\mathrm{max}}f < 0.25$ Oe/ns~\cite{hergt2007magnetic, dutz2013magnetic}, with frequency $f$ of a sinusoidal field of amplitude $H_{\rm max}$. (It is noteworthy that safe higher thresholds have been reported for particular types of cancerous tissue~\cite{albarqi2019biocompatible, simeonidis2016situ}.) As in I,  all of the dynamic hysteresis loops reported in the present study are performed at $T=310$~K, and we use SR=25~Oe/ns and $\alpha$=10.  This combination of SR and $\alpha$ is equivalent to the hyperthermia-relevant SR=0.25~Oe/ns and $\alpha$=0.1 for magnetite NPs.  This method of increasing $\alpha$ to simulate an effectively slower SR provides significant computational speed-up~\cite{BehCoarse-graining2020}.

The nanorod that we simulate has dimensions $8a_0 \times 24a_0 \times 56 a_0$ (with volume $V_{\rm rod}=6350.0$ nm$^3$), with its longest edge along the $z$ axis. The rod is made up of N$_b$ cubic cells with side length $a_b=ba_0$ ($b=1$, 2, 4, 8) while the volume of the rod is fixed for all simulations. A rod is composed of 10752 cells when the smallest cell (b=1) is used, and employing larger cells reduces the number of cells dramatically, as $N_b=10752/b^3$, to 1344, 168 and 21 for $b =2$, 4 and 8, respectively. Ultimately, RG scaling enables the description of a rod as a block, corresponding to $b=22$ ($\sqrt[3]{8\times24\times56}$), with a single magnetization vector with essentially the same hysteresis loop as obtained with the smallest cell size, even with magnetostatic interactions included. The impact of coarse-graining on loops is then examined for collections of nanorods that form nanoparticles as a foundation for simulating groups of NPs; see Fig.~\ref{fig:micromagnetics}.
 
In calculating hysteresis loops for any cell size, we apply an external magnetic field along the $z$ axis of $H(b) = H_{\rm max} \sin{(2 \pi f t)}$.
When uniaxial anisotropy is present, anisotropy directions for different cells within a nanorod are given by small random angles from the long axis of the rod (usually the $z$-axis) drawn from a normal distribution with a standard deviation of 5$^{\circ}$, i.e., anisotropy is along the long axis but with a small dispersion to imitate lattice disorder~\cite{plumer2010micromagnetic, serantes2014multiplying}.
$M(b)$ is the $z$ component of the magnetization, which we calculate by averaging over 90 to 100 independent simulations (averaging at each value of the field).  We report either $M(b)$ or its normalized form $m_H = M(b)/M_s$. 
At the beginning of a loop calculation, magnetic moments are randomized and $M(b)$ is approximately zero.  For the first quarter period, $H(b)$ goes from 0 to $H_{\mathrm max}$, and we report results for the subsequent period.

The error bars for the coercive field $H_c$ are calculated as one standard error above and below its mean value, obtained by considering the standard deviation in $H_c$ over the simulation ensemble used for each loop calculation. 

\section{\label{sec:RG} Coarse-graining and demagnetization}

\begin{figure*}\centering
    \includegraphics[width=0.45\textwidth]{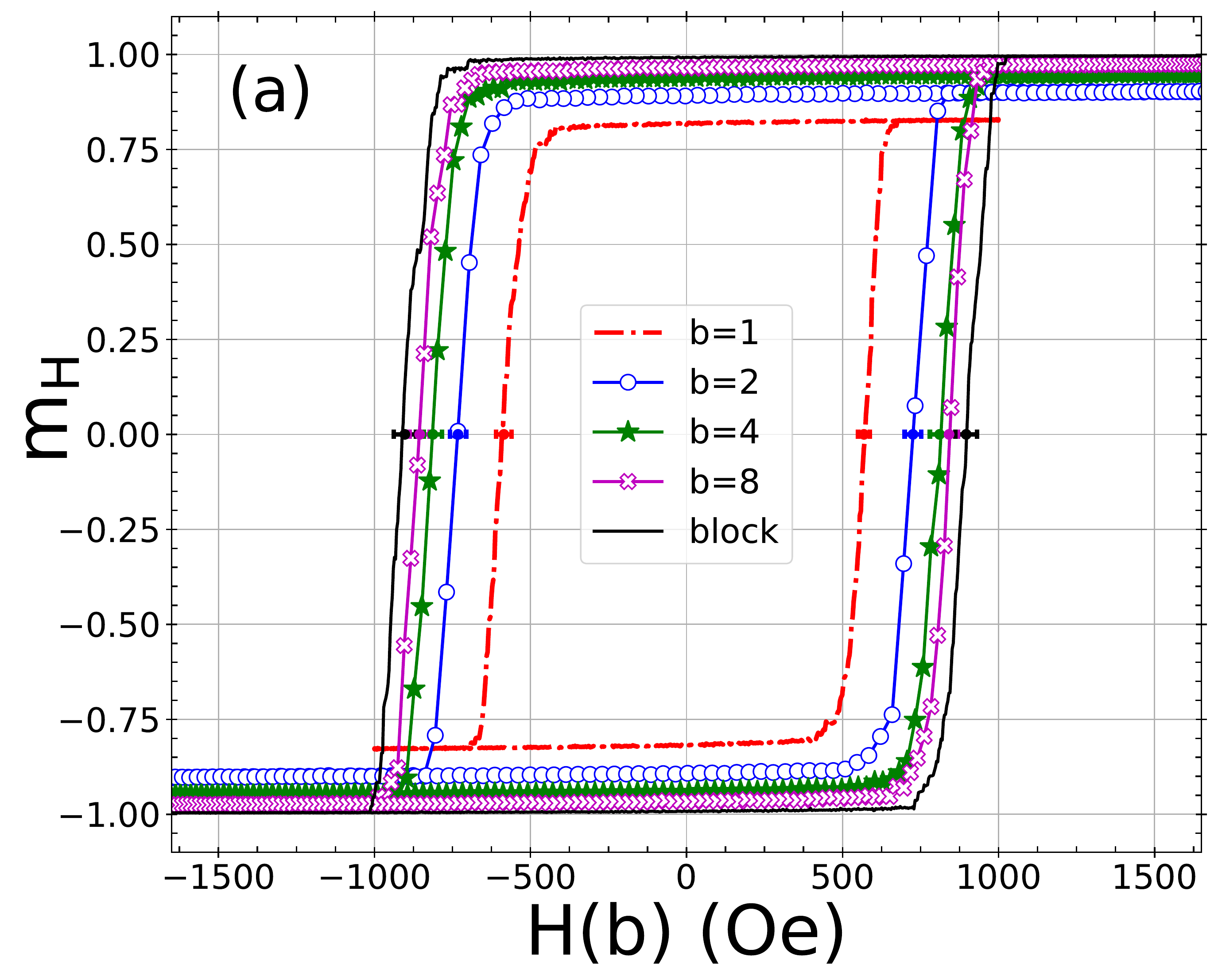}
    \includegraphics[width=0.45\textwidth]{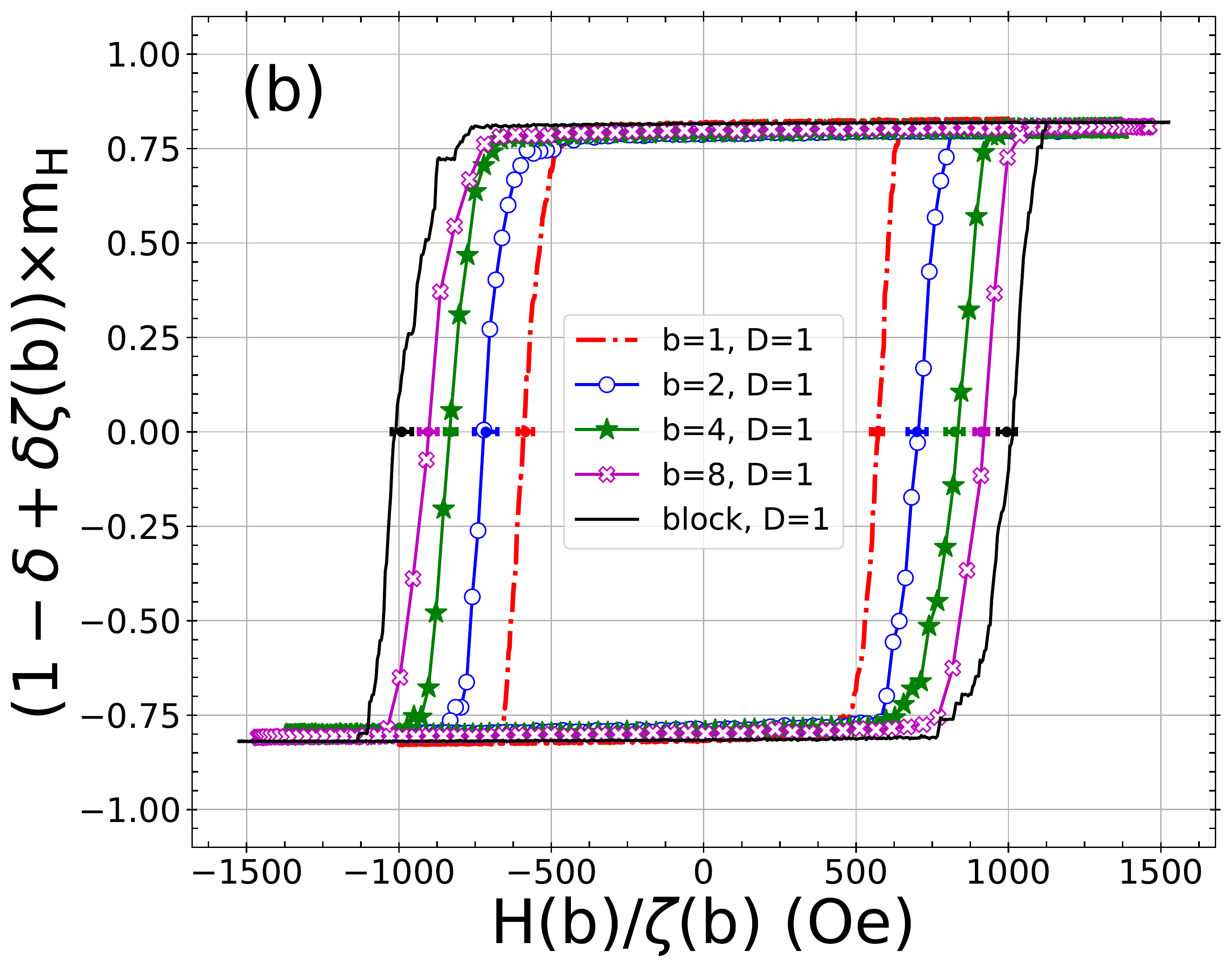}
    \includegraphics[width=0.45\textwidth]{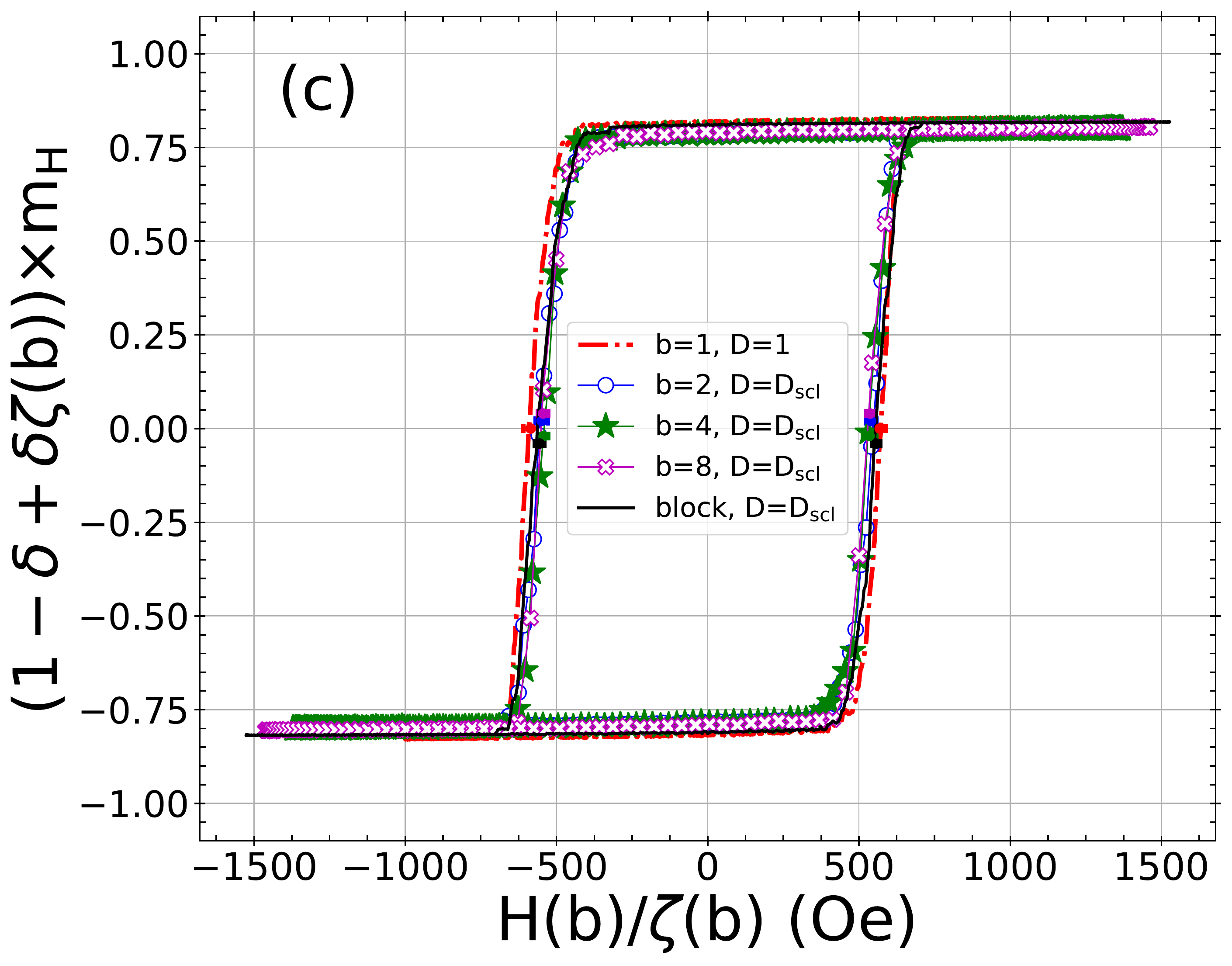}
    \caption{(a) Rod hysteresis loops when none of the magnetic parameters are scaled. (b) Scaling based on our modified Grinstein-Koch RG method~\cite{grinstein2003coarse,BehCoarse-graining2020} ($\delta\simeq$0.511) as in Eqs (1) - (5), but with no scaling of  magnetostatic interactions. (c) Scaling magnetostatic energy is included with the factor $D_{\mathrm{scl}}=\zeta(b)^{3}$. }
    \label{fig:RNG}
\end{figure*}

A few different approaches to scaling magnetic parameters such as $K$ and $A$ with simulation cell size have been proposed in the literature~\cite{grinstein2003coarse, CoarseGrainingFengVisscher, kirschner2005cell, kirschner2006relaxation}. 
As presented in I, we follow a modified version of the RG approach of Grinstein and Koch~\cite{grinstein2003coarse}, which results in a set of equations for the magnetization, exchange stiffness, applied field, and anisotropy constant,
\begin{eqnarray}
    M_0 &=& \delta \zeta(b) M(b) + (1-\delta) M(b) \label{eq:corrM0}\\
    A(b) &=& \zeta(b) \times A_0 \label{eq:RNG_scalingA}\\
    H(b) &=& \zeta(b) \times H_0 \label{eq:RNG_scalingH}\\
    K(b) &=& \zeta(b)^3 \times K_0 \label{eq:RNG_scalingK}
\end{eqnarray}
where,
\begin{equation}\label{eq:RNG_zeta}
     \zeta(b)=t/b+1-t, \qquad t=T/T_c, 
\end{equation}
$A_0$, $K_0$, $H_0$ and $M_0$ are the quantities for simulations using cell size $a_0$, $T_c$ is the critical temperature, and the quantities $A(b)$, $K(b)$, $H(b)$ and $M(b)$ are those for a simulation where the cell size is $a_b=ba_0$.  The phenomenological parameter $\delta=0.511$ was determined in I from the $T$ dependence of $M$ for our nanorods. 
In  the  present  work,  we  propose  and  test  a  scaling  for  magnetostatic interactions, not included in the work of Grinstein-Koch or in I.

As a first step in determining a scaling for magnetostatic interactions,
we calculate a reference hysteresis loop for $b=1$ for a maghemite nanorod by running simulations using $A_0$ and $K_0$ for the exchange and uniaxial anisotropy parameters, respectively, and include magnetostatic interactions. Results are given by the red curve in all panels of Fig.~\ref{fig:RNG}.
We then carry out loop simulations with cell sizes $b a_0$, for $b=2$, 4, and 8.  For $b=22$, the dimensions of the single cell are those of the nanorod itself. 
For these simulations, we use unrenormalized exchange and anisotropy parameters $A(b)=A_0$, $K(b)=K_0$, and again include magnetostatic interactions. 
The loops resulting from these non-scaled simulations are plotted in Fig.~\ref{fig:RNG}a, showing a very significant increase in loop size as cell size increases.

We repeat the loop calculations for $b>1$ using values of $A(b)$ and $K(b)$ from Eqs.~\ref{eq:RNG_scalingA} and \ref{eq:RNG_scalingK}, respectively, and with $M$ and $H$ scaled via Eqs.~\ref{eq:corrM0} and \ref{eq:RNG_scalingH}, so that we plot $m_H=M_0/M_s=(\delta \zeta(b)+1 -\delta)M(b)/M_s$ as a function of $H_0=H(b)/\zeta(b)$, and again we include full magnetostatic interactions. The resulting hysteresis loops are different for different $b$, with coercivity increasing with cell size, as shown in Fig.~\ref{fig:RNG}b.



From the above results, it is clear that magnetostatic interactions need to be scaled as cell size changes.
Looking at the energy terms in the Hamiltonian (see Appendix~\ref{app:time_stp}) and noting that the exchange energy ($aA\sum m_i.m_j$) is proportional to the cell length and $A$ is scaled with $\zeta(b)$, whereas the magnetocrystalline anisotropy energy ($K_uv\sin^2(m_i.u)$) is proportional to the cell volume and $K_u$ is scaled with $\zeta^3(b)$, we propose a $\zeta^3(b)$ scaling for the demagnetization energy, which is also proportional to the cell volume.
The magnetostatic energy is $\mu_0 v M_s^2 m\cdot N\cdot m/2$, where the demagnetization tensor $N$ is determined by the geometry of the system.
We repeat the loop calculations for $b >1$, again using RG scaling for $A$, $K$, $M$ and $H$, but now multiply magnetostatic energies and torques by $\zeta(b)^3$. As can be seen in Fig.~\ref{fig:RNG}c the collapse of the data is reasonably good.
The loop areas for $b=1$, 2, 4, 8 and block simulations are 1881, 1706, 1691, 1703, 1800 Oe,
respectively. The smallest loop area (for $b=8$) is 9\% smaller than the area for $b=1$. We note that comparing the above hysteresis loops with a system without magnetostatic interactions (Fig. 2c in I) supports a result from Mehdaoui~et al.~\cite{mehdaoui2013increase}, namely, that including magnetostatic interactions increases the squareness of the loops.

To accomplish the scaling of magnetostatic interactions when using OOMMF, we take the approach of scaling $M_s$, while ensuring that all other terms in the effective field remain unchanged. The magnetostatic energy is proportional to $M_s^2$. Therefore, multiplying $M_s$ by $\zeta(b)^{3/2}$ results in the desired scaling of magnetostatic interactions with $\zeta(b)^3$. At the same time, scaling $M_s$ changes the non-magnetostatic contributions to the effective field entering the LLG calculations, namely the exchange, anisotropy and thermal contributions. 
We must therefore introduce addition scaling to preserve $\mathbf {H}_{\mathrm{eff}}=\mathbf{H}_{\mathrm{exch}}+\mathbf{H}_{\mathrm{anis}}+\mathbf{H}_{\mathrm{ext}}+\mathbf{H}_{\mathrm{thermal}}$ invariant to changes in $M_s$. 
Thus, when changing program input $M_s$ to $M_s\zeta(b)^{3/2}$, we must additionally change $A$ to $A\zeta(b)^3$, $K$ to $K\zeta(b)^{3/2}$ and $T$ to $T\zeta(b)^{3/2}$ in order to keep field strengths 
$H_{\mathrm{exch}}=2A/\mu_0 a M_s^2$, 
$H_{\mathrm{anis}}=2K/\mu_0 M_s$, and 
$H_{\rm thermal}= \left[2\alpha k_B T/(\gamma \mu_0 M_s V \Delta t)\right]^{1/2}$ unaltered. The end result is that in order to carry out a simulation at $b>1$ and temperature $T_0$, we first calculate $\zeta=\zeta(T_0, b)$, and the set program inputs to
$M_s = M_{s0} \zeta^{3/2}$,
$A = A_0 \zeta^6$, $K=K_0 \zeta^{9/2}$, and $T=T_0\zeta^{3/2}$. The external field $H(b)$ is unchanged. This recipe combines the RG scaling of $A$ and $K$ with appropriate scaling of magnetostatics, and yields $M(b)$.


The next step is to model the collective effect of the magnetocrystalline anisotropy, exchange and magnetostatic interactions of a rod  with a single magnetization (macrospin) subject to uniaxial anisotropy.  
This step is justified by the rather good agreement in the MH loops between the fine grain simulation ($b=1$), and the single block case ($b=22$), for which a single magnetization represents the entire rod and no explicit exchange interactions are present. 
This macrospin description is know as the Stoner-Wohlfarth (SW) model, and the Hamiltonian is,
\begin{equation}
    \begin{split}
        &\mathcal{H}=  \mathcal{H}_{\mathrm{anisotropy}} + \mathcal{H}_{\mathrm{Zeeman}},\\
    & \mathcal{H}_{\mathrm{anisotropy}}=-K_{\mathrm{eff}}v (\mathbf{m}\cdot\mathbf{u})^2,\\
    & \mathcal{H}_{\mathrm{Zeeman}} = -\mu_0M^{\rm eff}_sv (\mathbf{m}\cdot \mathbf{H}),
    \end{split}
\end{equation}
where the uniaxial anisotropy has energy density $K_{\mathrm{eff}}$ with its axis along $\mathbf{u}$, and the single magnetization vector has direction $\mathbf{m}$ and magnitude $M^{\rm eff}_s$.  $K_{\mathrm{eff}}$ and $M^{\rm eff}_s$ arise from the combined effects of self-demagnetization, magnetocrystalline anisotropy, exchange, and temperature.  For the macrospin model of the nanorod $v=V_{\rm rod}$.
%
%
$\mu_0$ is the permeability of free space and $\mathbf{H}$ is the externally applied field.
This SW macrospin model may be useful for simulating a group of nanorods in solution, for example, and it is understood that interactions between rods include magnetostatic interactions, perhaps in the dipole approximation. This macrospin description differs from the $b=22$ block model in that, first, the self-magnetostatic interaction is accounted for by the effective uniaxial anisotropy, and second, there is no need to worry about the procedures to implement RG and magnetostatic scaling.

\begin{figure}
    \centering
    \includegraphics[width=0.49\textwidth]{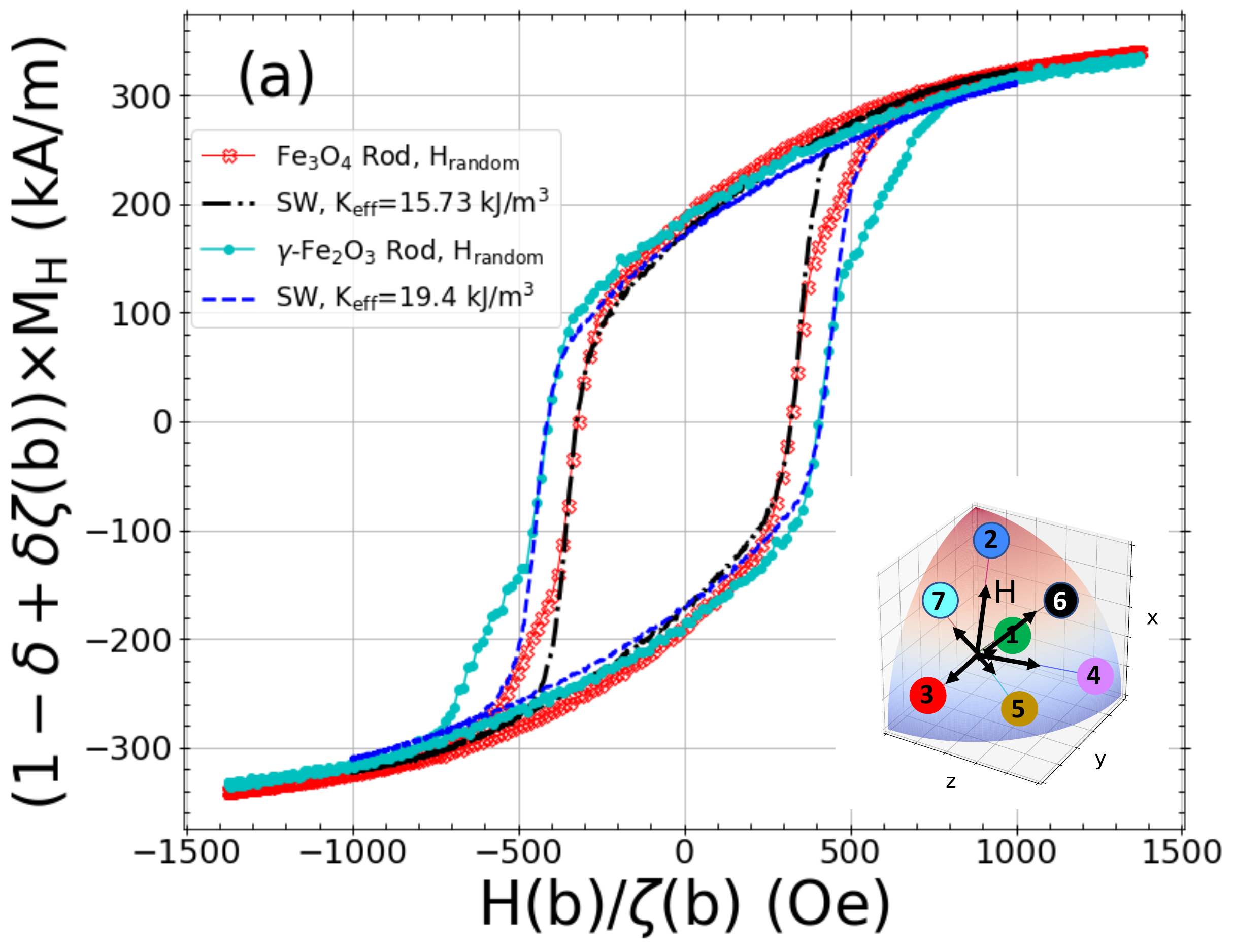}
    \includegraphics[width=0.49\textwidth]{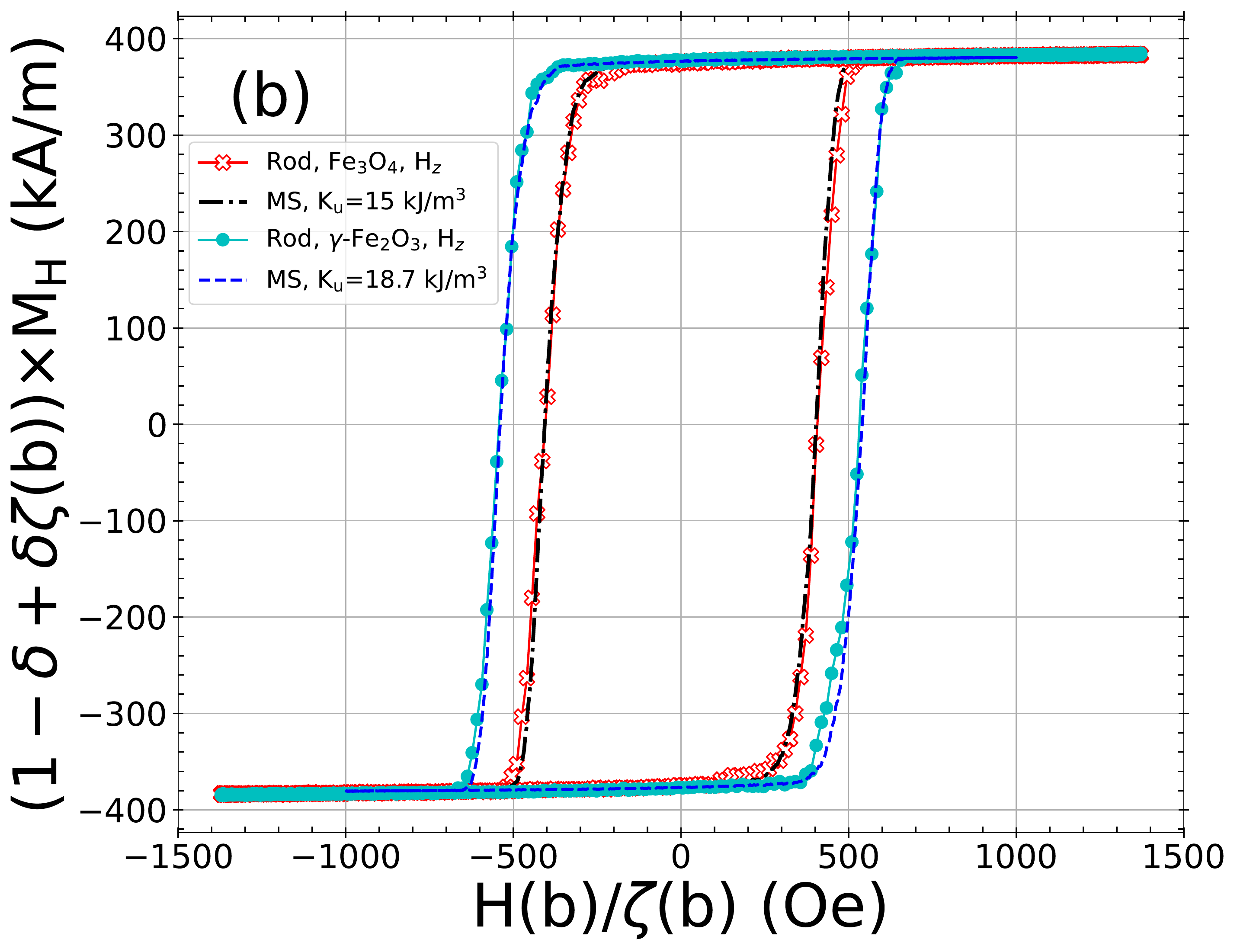}
    \caption{Comparison of macrospin models with 
     magnetite and maghemite nanorods for (a) rotationally averaged external field -- SW refers to the macrospin in this case, and (b) external field along the $z$ axis -- MS refers to the macrospin in this case. 
     }
    \label{fig:RodSW}
\end{figure}

To find the appropriate parameters to model the nanorod as a SW-macrospin at 310~K, we calculate the hysteresis loop of the nanorods modelled using $b=4$, averaging over directions between field and long nanorod axis.  Given the symmetry of the rod, it is sufficient to integrate directions over a spherical octant, and, following the numerical algorithm presented in Ref.~\cite{bavzant1986efficient}, we employ a seven-point integration scheme, with directions shown in the inset of Fig.~\ref{fig:RodSW}a.   We also calculate the directionally averaged loop for a SW particle at 310~K by simulating 1000 particles with random orientations (uniformly over a shpere), and then scaling the parameters of the SW particle to match $H_c$ and remanent magnetization $M_r$ of the rod.
For a magetite rod ($K=0$), we find that $K_{\rm eff}=15.7$~kJ/m$^3$ and $M^{\rm eff}_s = 0.73 M_s = 350$~kA/m.  Results are plotted in Fig.~\ref{fig:RodSW}a. It is important to note that if one wished to plot $m_H$, one should normalize $M_H$ by $M_s$, rather than by $M^{\rm eff}_s$, in order to compare with nanorod loops.
For a maghemite rod ($K_0=10$~kJ/m$^3$), we find $K_{\rm eff}=19.4$~kJ/m$^3$ and $M^{\rm eff}_s = 0.73 M_s = 350$~kA/m.

From the loops  shown in Fig.~\ref{fig:RodSW}a, it is clear that the rod does not precisely follow the SW model.  This is because the magnetostatic interactions within the rod only approximately map to a single anisotropy axis.  In Fig.~\ref{fig:RodSW}b, we plot the MH loops for the $b=4$ approximation for the rod and the SW counterpart when the field is along the $z$ axis, i.e., along the anisotropy axis.  In this case, we find a smaller value of $K_{\rm eff}=15.0$~kJ/m$^3$ for magnetite, with $M_s^{\rm eff} = 0.8 M_s=384$~kA/m. 
This value of $K_{\rm eff}$ is smaller than the analytical result at $T=0$, $K^{T=0}_{\rm eff}=20.5$~kJ/m$^3$, which we obtain by following Refs.~\cite{cullity2011introduction, Newell1993generalization, fukushima1998volume, Aharoni1998}.
For maghemite, we obtain 
$K_{\rm eff}=18.7$~kJ/m$^3$ with $M_s^{\rm eff} = 0.80 M_s=384$~kA/m.  All effective parameters are summarized in Table~\ref{table:Keff}.

\section{Coarse-graning for multiple nanorods\label{multiple_nanorods}}
\begin{figure*}
    \centering
    \includegraphics[width=0.49\textwidth]{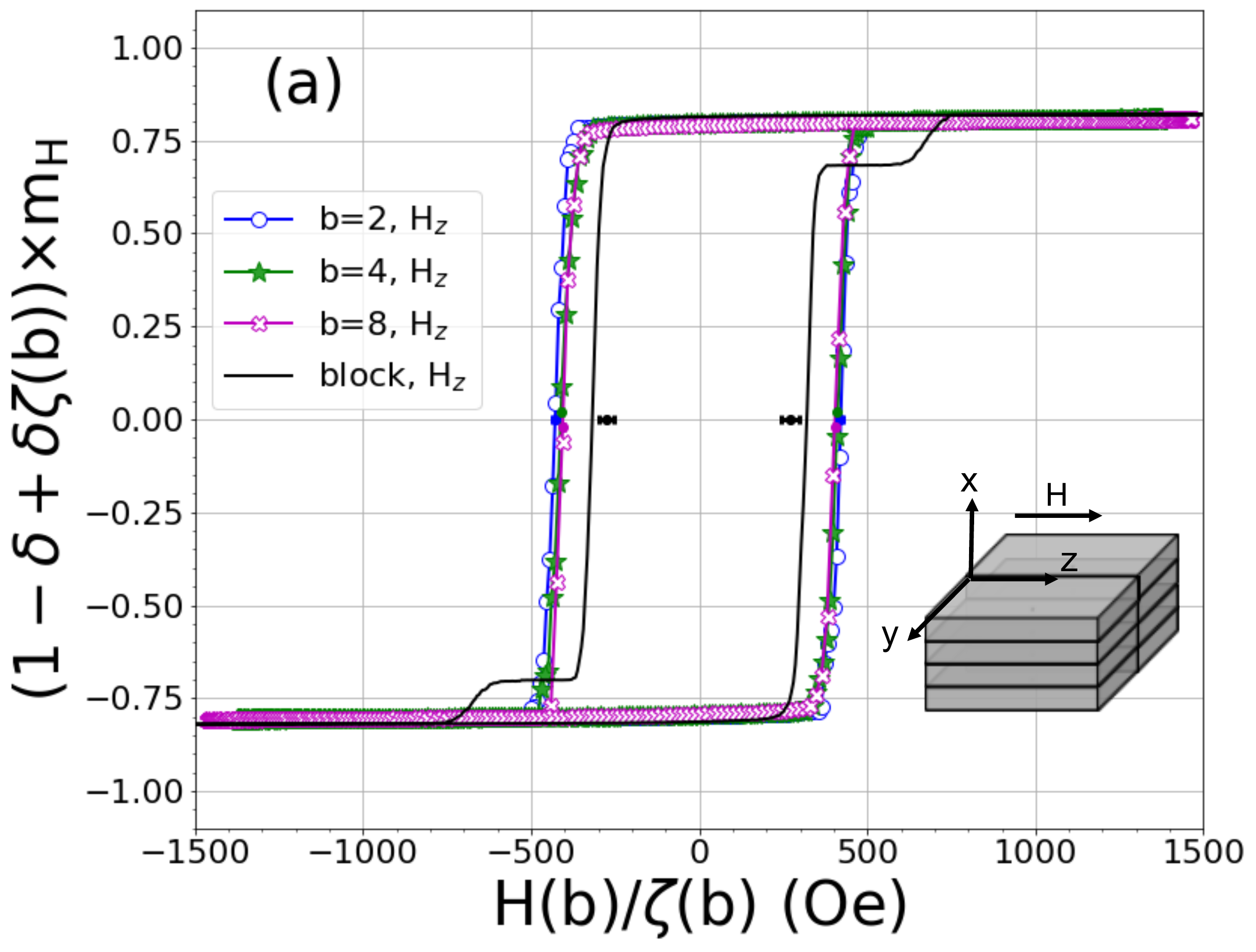}
    \includegraphics[width=0.49\textwidth]{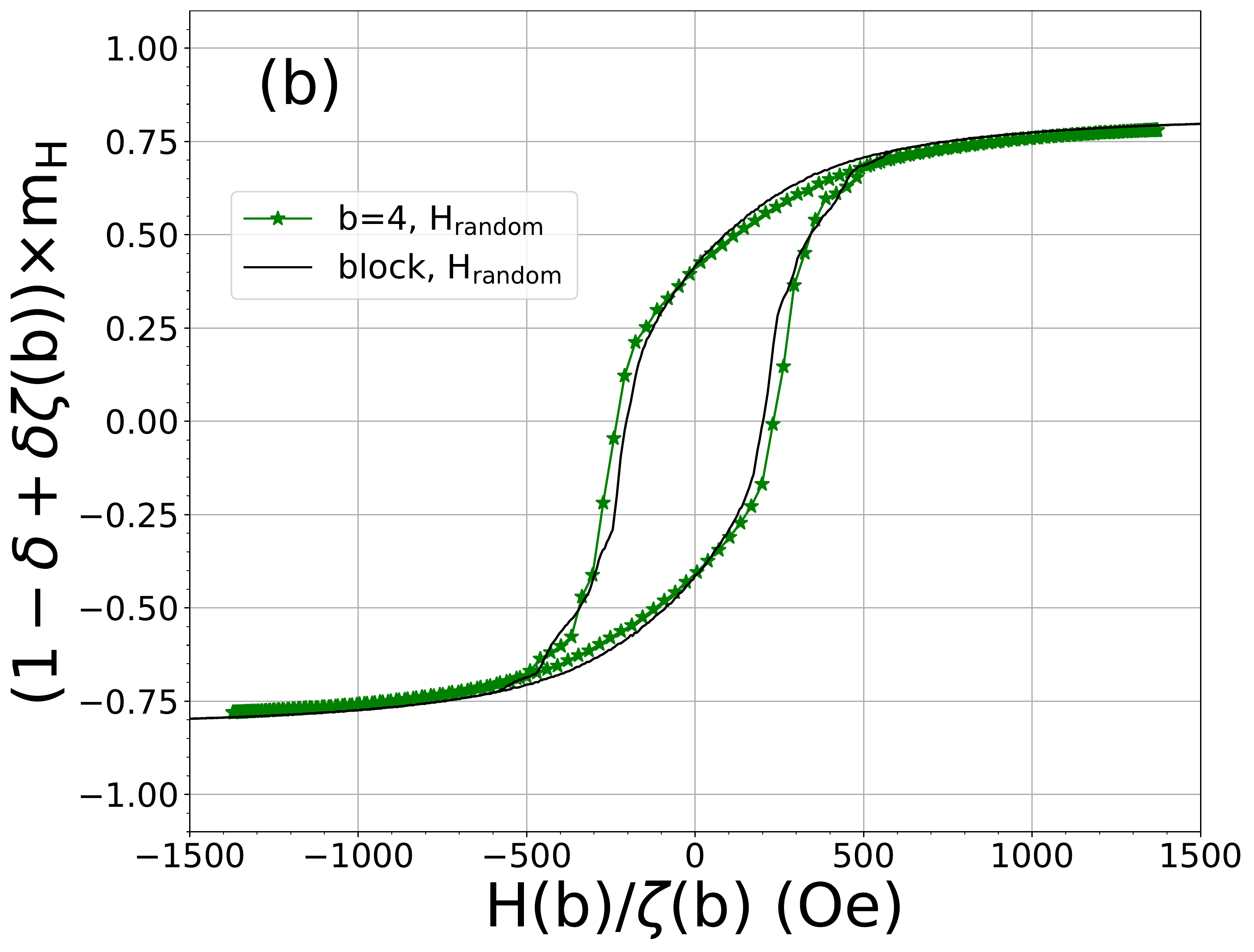}
    \caption{Scaling applied to a bundle of 8 nanorods with inter-rod exchange $A_{\rm r-r}=0.5 A(b)$. (a)  Loops corresponding to simulation cells of length $a_b=b a_0$ for $b=2$, 4, 8 and 22 (block) for a field applied along the $z$ axis. (b) Loops with a rotationally averaged field for nanorods modelled with $b=4$ and 22 (block). }
    \label{fig:8Rod_scaling}
\end{figure*}

As our goal is to simulate magnetic nanoparticles made of nanorods, we test the proposed scaling method for a collection of eight maghemite nanorods in two stacks of four as shown in the inset of Fig.~\ref{fig:8Rod_scaling}.
Simulations include magnetostatics, intrarod [$A(b)$] and inter-rod ($A_{\rm r-r}$) exchange interactions at half strength [$A_{\rm r-r}=0.5 A(b)$], magnetocrystalline uniaxial anisotropy along the rod's long axis and a sinusoidal field applied along the $z$ axis.

Simulated MH loops for the eight-rod bundle show good agreement for $b=2$, 4 and 8, whereas the loop is significantly different for a  bundle of eight blocks ($b=22$) as shown in Fig.~\ref{fig:8Rod_scaling}a. Clearly, modelling the nanorod as a block with a single magnetization does not allow portions of a nanorod to flip independently of the rest of the rod, and hence the shoulder regions of the loop in particular are susceptible to unphysical behaviour. 
Thus, magnetostatic interactions limit the present prescription for course-graining in the case of bundled nanorods.

We expand our exploration by comparing the average MH hysteresis loop of this group of nanorods when the applied field is  rotationally averaged. 
Interestingly, averaging over field directions masks the discrepancy between $b=4$ and the block approximation, as shown in Fig.~\ref{fig:8Rod_scaling}b.  We conclude that $b=4$ is a reasonable level of coarse-graining for the investigation of multiple-rod configurations in the remainder of the present work.

\section{\label{sec:2rods}Various 2-rod setups}

In this section we qualitatively explore the effects of magnetostatic and exchange interactions for three different arrangements of two magnetite nanorods, providing some insight on their  effects on the magnetization alignment for bundled nanorods. We use RG scaling with $b=4$, and, for this section only, we do not  carry out the scaling of magnetostatic interactions, and simply use $M_s$ with no alteration in determining effective fields and energies.  We are simply interested in general effects of the interplay between magnetostatics and inter-rod exchange.


\begin{figure*}
    \centering
    \includegraphics[width=0.45\textwidth]{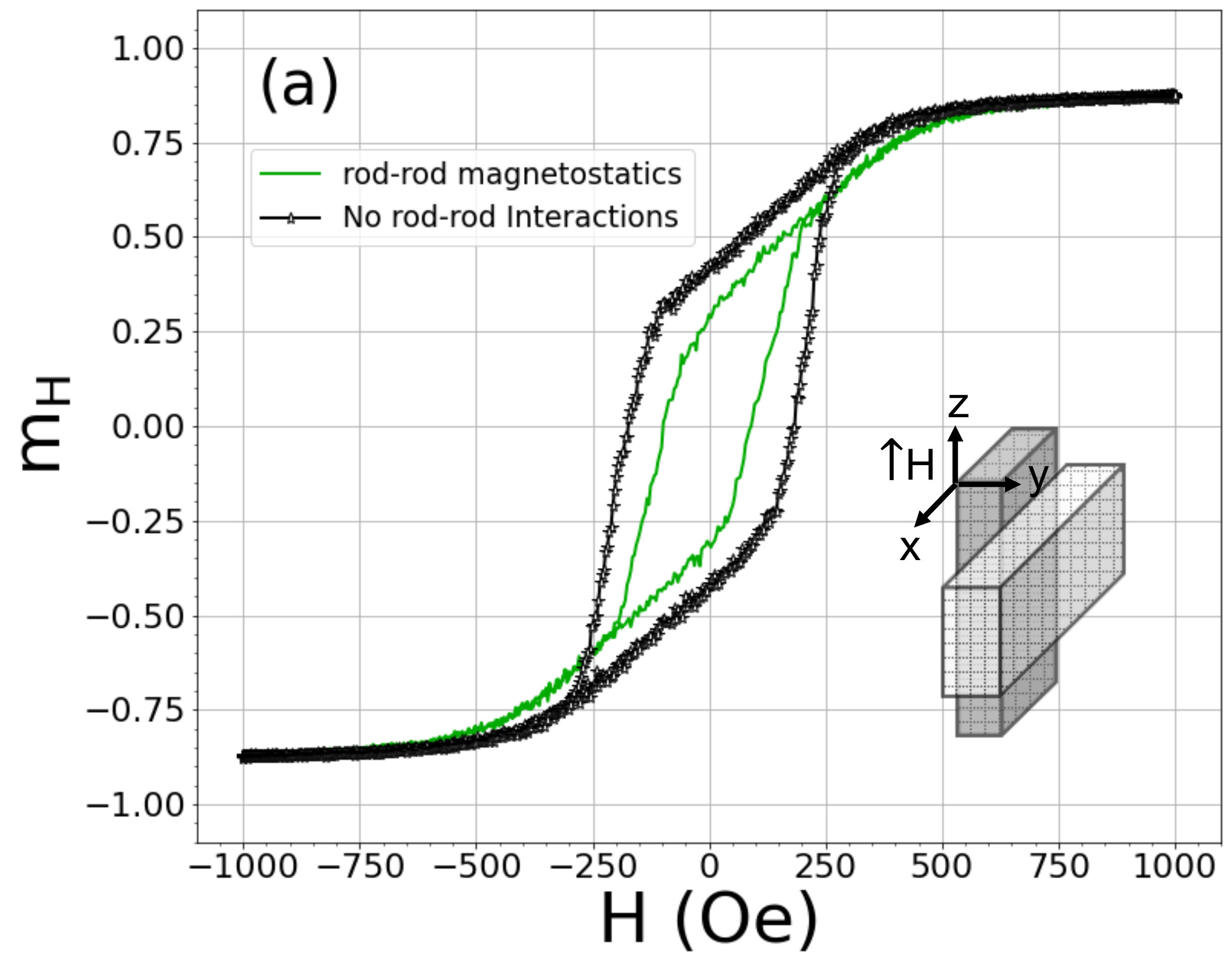}
    \includegraphics[width=0.45\textwidth]{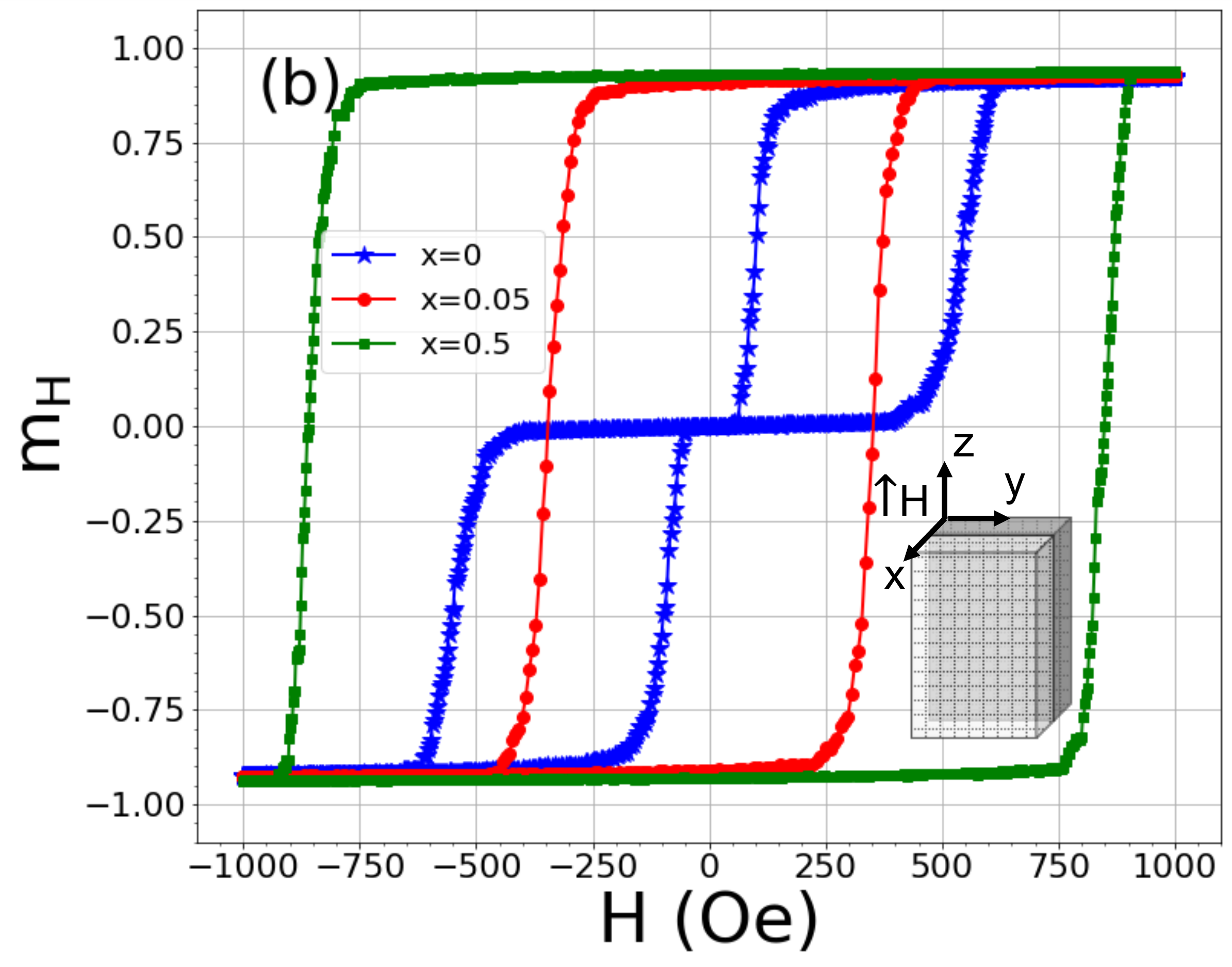}
    \includegraphics[width=0.45\textwidth]{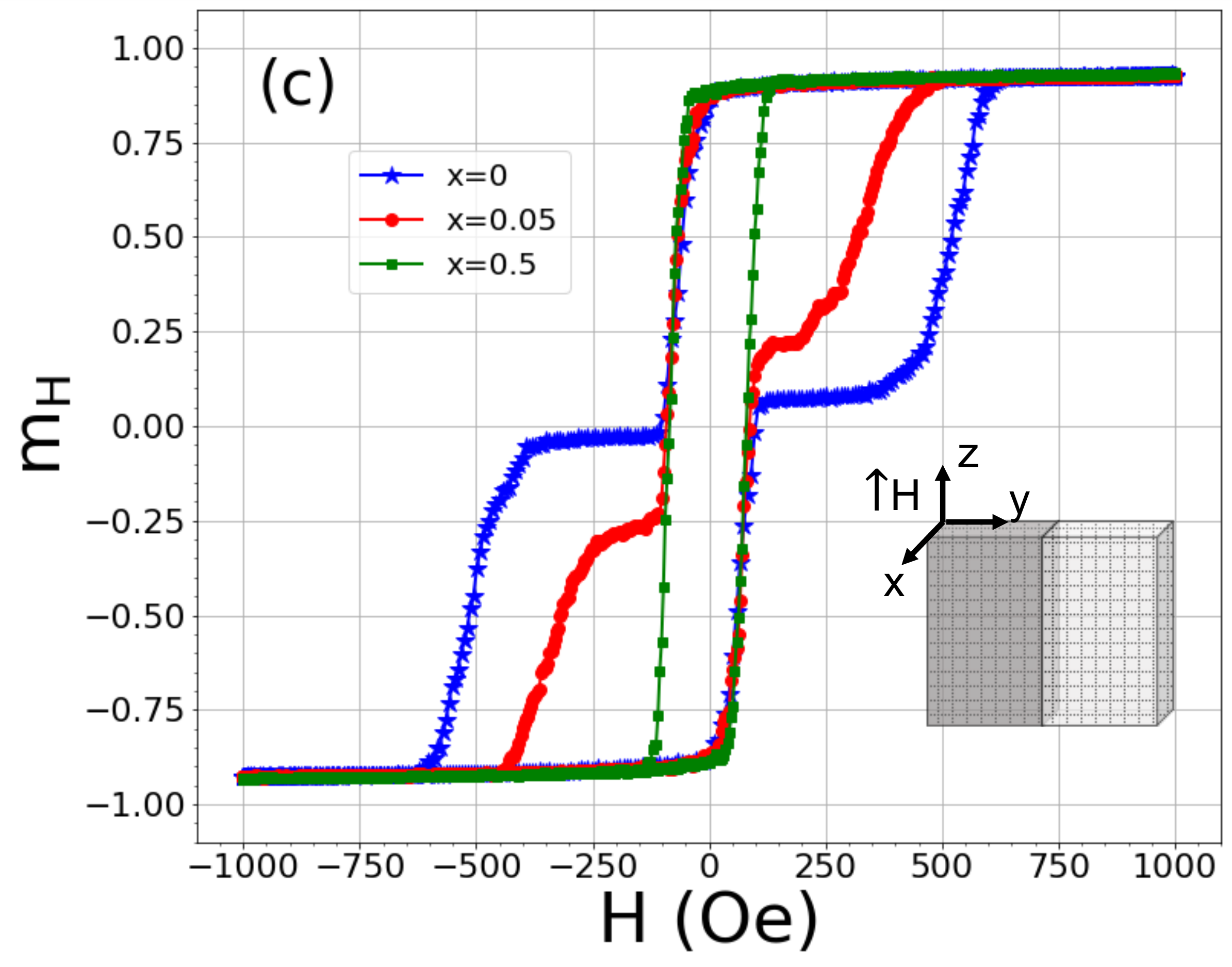}
    \caption{(a) Effect of inter-rod magnetostatic interactions.  The black loop (stars) is for two perpendicular noninteracting nanorods (with neither exchange, nor magnetostatics between rods) and the green loop is for nanorods interacting magnetostatically only. In panel (b) nanorods interact magnetostatically and inter-rod exchange is $A_{\mathrm{r-r}}=xA(b)$, with $x = 0$ for the blue curve (stars), 0.05 for the red curve (circles) and 0.5 for the green curve (squares). The two parallel nanorods are in contact with their largest faces and the center-to-center distance is 6.7~nm. Panel (c), as in (b), except a smaller face is shared, and center-to-center distance is 20~nm. In this case, increasing $x$ does not result in a larger loop area.}
    \label{fig:2rods}
\end{figure*}

In the first arrangement, we consider only the effect of magnetostatic interaction between rods. One nanorod is placed along the $z$ axis, and the other along $x$, with the $y$ axis passing through the nanorod centers, as shown in the inset of Fig.\ref{fig:2rods}a. The external field is along the $z$ axis. Within each rod, magnetostatics and exchange are present.  For the black curve in Fig.~\ref{fig:2rods}a, the rods do not interact: they are independent with $A_{\rm r-r}=0$ and with no magnetostatic interactions between cells belonging to different rods.  The loop, in fact, is just the average of two independent rods.  The green curve in the same plot shows the loop for the case where the two rods interact magnetostatically: magnetostatic interactions are calculated between all cells in the 2-rod system.  The hysteresis loop is smaller for the interacting case.  This negative effect of magnetostatics on loop area is in agreement with studies reported by Cabrera~et al.~\cite{cabrera2018dynamical} and Serantes~\cite{serantes2014multiplying}, wherein dipole interactions decrease the heating efficiency of  magnetic particles when the dipoles are not arranged in end-to-end chains.



Panels b and c of Fig.~\ref{fig:2rods} compare hysteresis loops, when inter-rod magnetostatic interactions are present, for three different inter-rod exchange strengths $A_{\mathrm{r-r}}=x A(b)$, with  $x$ = 0, 0.05, and 0.5.  Here, the nanorods are side-by-side with their long axes parallel. Fig.~\ref{fig:2rods}b considers the case of rods with their largest faces making contact (area of contact is 84 $a_4^2$), and Fig.~\ref{fig:2rods}c considers the case where the nanorods are making contact through their second largest faces (area of contact is 28 $a_4^2$). The centers of adjacent parallel nanorods are 6.7~nm and 20~nm apart in panels b and c, respectively. 
In general, increasing $x$ increases the magnetization alignment between the two nanorods, counteracting the anti-alignment induced by magnetostatics. In Fig.~\ref{fig:2rods}b, for $A_{\rm r-r}=0$ the magnetization of one rod flips before $H$ becomes negative.  For $A_{\rm r-r}=0.05 A(b)$ and 0.5$A(b)$, the magnetizations of the two rods are locked, and higher exchange strength results in wider hysteresis loops.
For the larger centre-to-centre separation (and therefore weaker inter-rod magnetostatic interactions) and smaller contact area presented in Fig.~\ref{fig:2rods}c, for $A_{\rm r-r}=0$, the magnetization of one of the rods flips before the other, but only after the $H$ becomes negative.  
At $A_{\rm r-r}=0.05 A(b)$, when the magnetization of one rod flips, it takes part of the second rod with it.  Only at $A_{\rm r-r}=0.5 A(b)$ do the magnetizations of both rods flip in unison. We note that for $b=4$, the exchange length is $\sqrt{\frac{2 \zeta(4) A_0}{\mu_0 M_s^2}} \approx 7.0$~nm, and therefore significantly smaller than the centre-to-centre distance. The perhaps counter-intuitive observation is that as $A_{\rm r-r}$ increases, the loop area decreases.  We conclude that the pairing of inter-rod exchange and magnetostatics can lead to complex magnetization dynamics within nanorod composites, and therefore counter-intuitive impacts of inter-rod exchange on heating efficiency.


In all 2-rod cases considered, we explicitly place the rods side-by-side and not end-to-end. Thus, we do not consider chain formation~\cite{torche2020thermodynamics}, which should enhance hysteresis, but rather the tendency of magnetostatics to cause anti-alignment of neightbouring nanorod magnetic moments.  We note that the larger centre-to-centre distance considered in Fig.~\ref{fig:2rods}c means that the anti-aligning effects of magnetostatics is weaker, and so perhaps it is not surprsing to see a larger loop area than in  Fig.~\ref{fig:2rods}b in the $A_{\rm r-r}=0$ case.


\section{Nanoparticles\label{sec:np}}

\begin{figure*}
    \centering
    \includegraphics[width=0.47\textwidth]{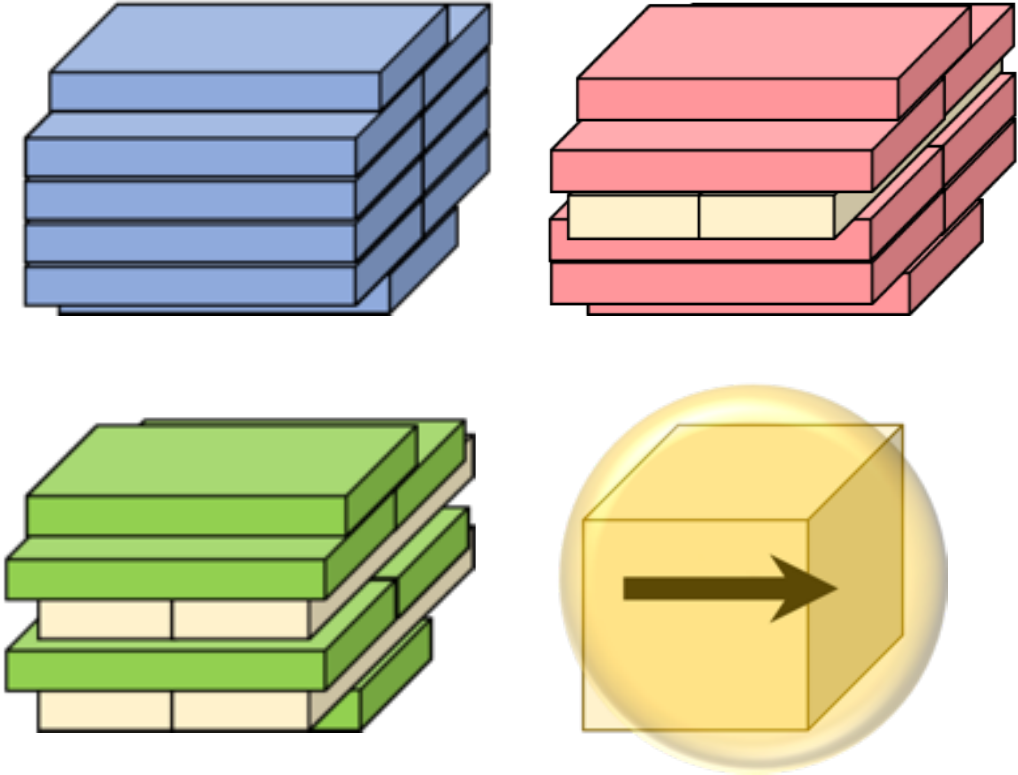}
    \includegraphics[width=0.52\textwidth]{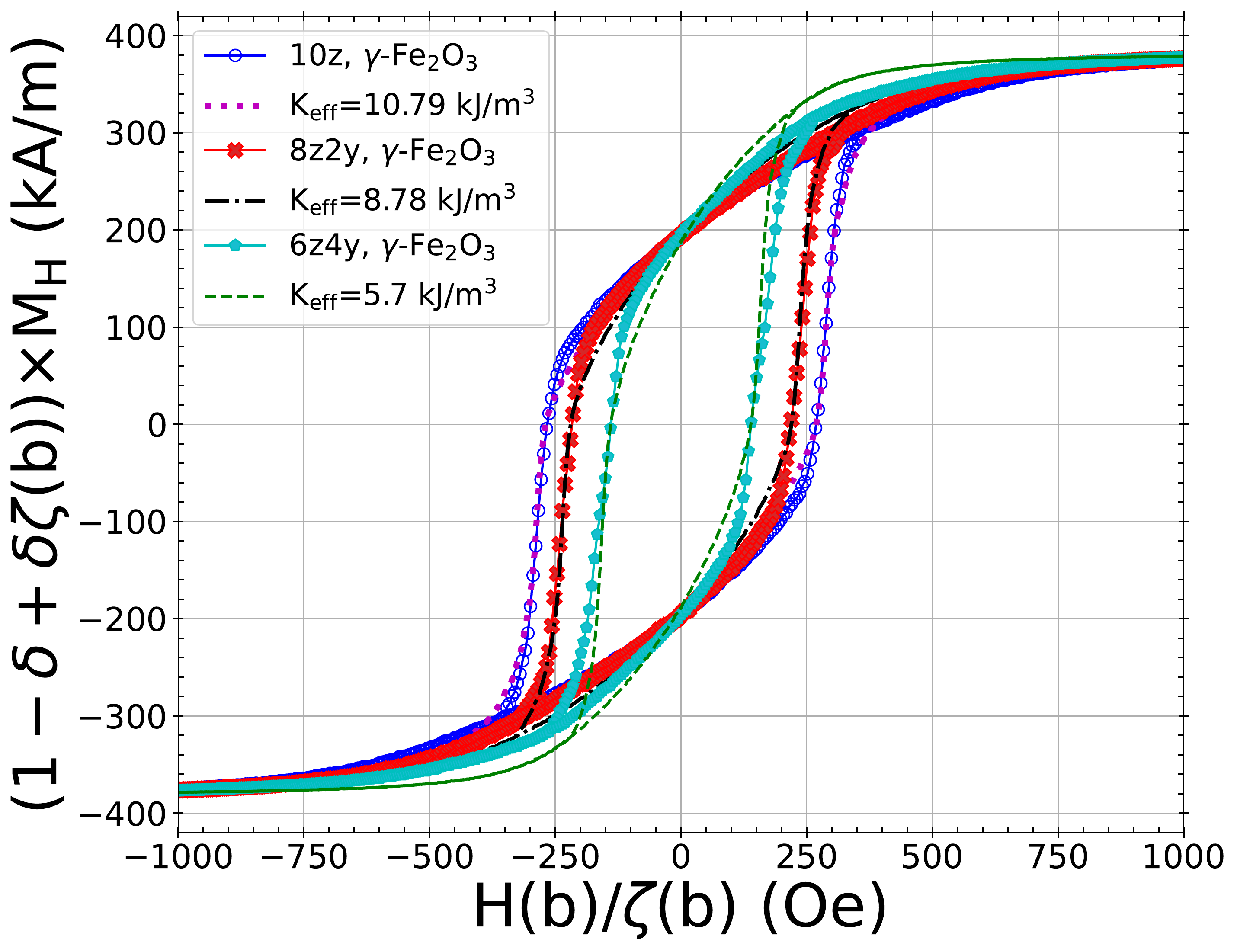}
    \caption{ 
    Three different NPs, $10z$, $8z2y$ and $6z4y$, each assembled from 10 maghemite nanorods.
    The right graph shows the NP hysteresis loops for rotationally averaged field (solid curves with symbols), and loops for their equivalent macrospins with the same $M_r$ and $H_c$ (dashed curves). 
    Macrospins equivalents to each NP have $K_{\rm eff}= 5.7$, 8.78 and 10.79 kJ/m$^3$ for $6z4y$, $8z2y$ and $10z$ NPs, respectively, and $M^{\rm eff}_s = 382$~kA/m.}
    \label{fig:NP_SW}
\end{figure*}

Our basic model of nanoparticles composed of nanorods is inspired from the experimental study  by Dennis~et al.~\cite{dennis2009nearly}. There are, however, no data on how nanorods are packed within a nanoparticle, and two extreme possible assemblies are a totally ordered stack of nanorods and a random cluster of nanorods~\cite{pearce2013magnetic}. Among various possible arrangements, we choose three assemblies containing 10 maghemite ($K_0=10$ kJ/m$^3$) nanorods, one with all the nanorods along the $z$ axis (which we label $10z$), another one with 8 along the $z$ axis and 2 along the $y$ axis ($8z2y$) and a third arrangement with 6 nanorods along  $z$ and 4 along $y$ ($6z4y$), as shown in Fig.~\ref{fig:NP_SW}a.  With these three choices, we mimic some degree of disorder by varying the degree of rod alignment.   To compare the heating efficiency of these constructions with the experimental results, we calculate the rotationally averaged hysteresis loop, coarse-graining the rods at the $b=4$ level (including magnetostatic scaling) and assuming $A_{\rm r-r}=0.5 A(b)$. As expected, assemblies with more parallel nanorod arrangement exhibit wider hysteresis loops, as shown in Fig.~\ref{fig:NP_SW}b, which leads to higher heating efficiency.

The next step in simplifying the simulation of NPs is to find the magnetic parameters of a SW macrospin that gives the most similar MH hysteresis loops (the same $M_r$ and $H_c$) to nanoparticles of the same volume. This level of modelling enables the description of a complex nanoparticle made of nanorods with a single macrospin and replacing all the magnetostatic and exchange interactions inside the NP with an effective uniaxial anisotropy of the macrospin.
The resulting fits, made by adjusting $K_{\rm eff}$ and $M_s^{\rm eff}$, are shown in Fig.~\ref{fig:NP_SW}b, and the effective uniaxial anisotropy for the three maghemite nanoparticle models $10z$, $8z2y$ and $6z4y$  are 10.79, 8.78 and 5.7~kJ/m$^3$, respectively, with effective saturation magnetization equal to $0.795 M_s = 382$~kA/m for all three models. 
Effective parameters for maghemite and magnetite nanoparticles are given in Tab.~\ref{table:Keff}.


\begin{figure}
    \includegraphics[width=0.49\textwidth]{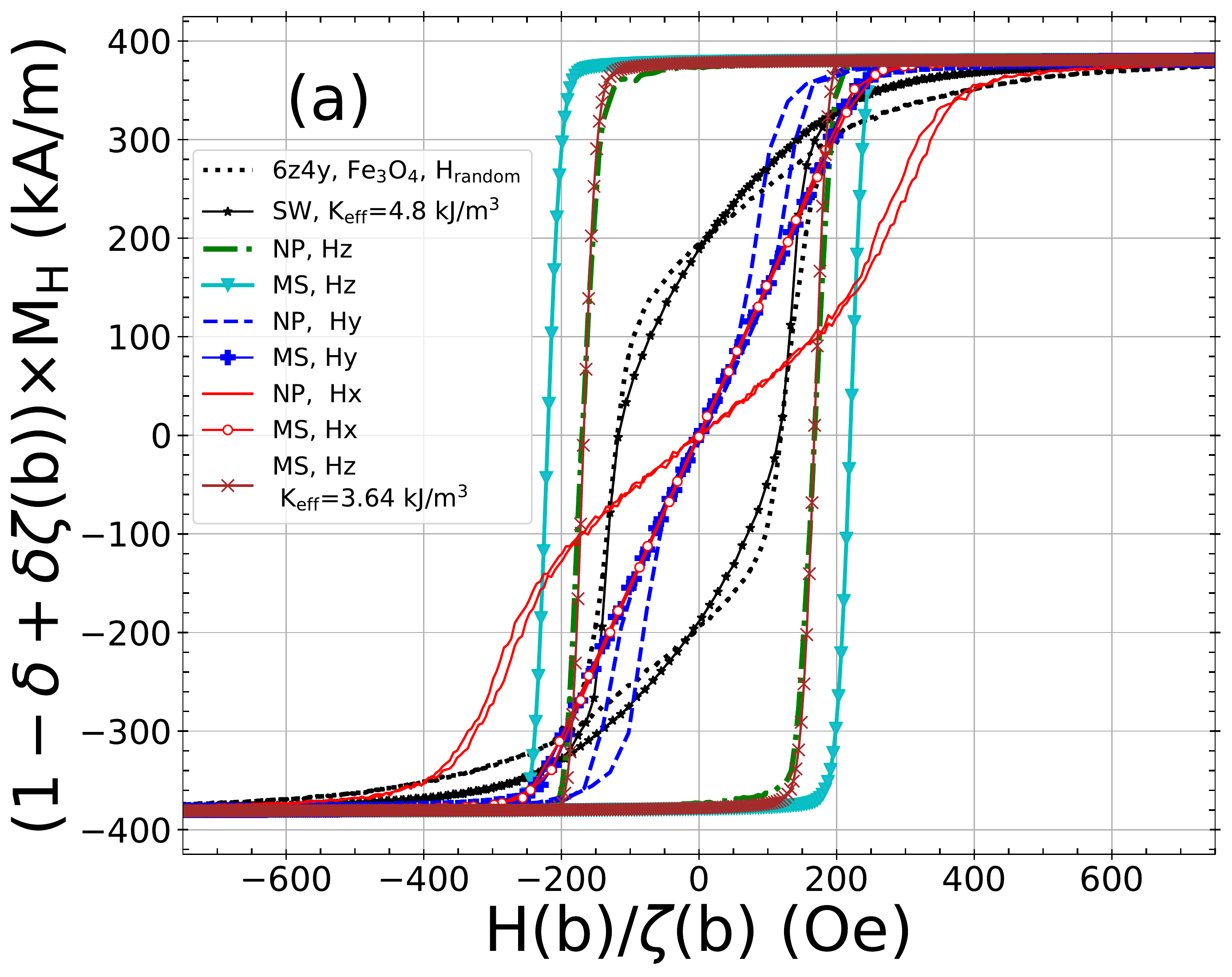}
	\includegraphics[width=0.49\textwidth]{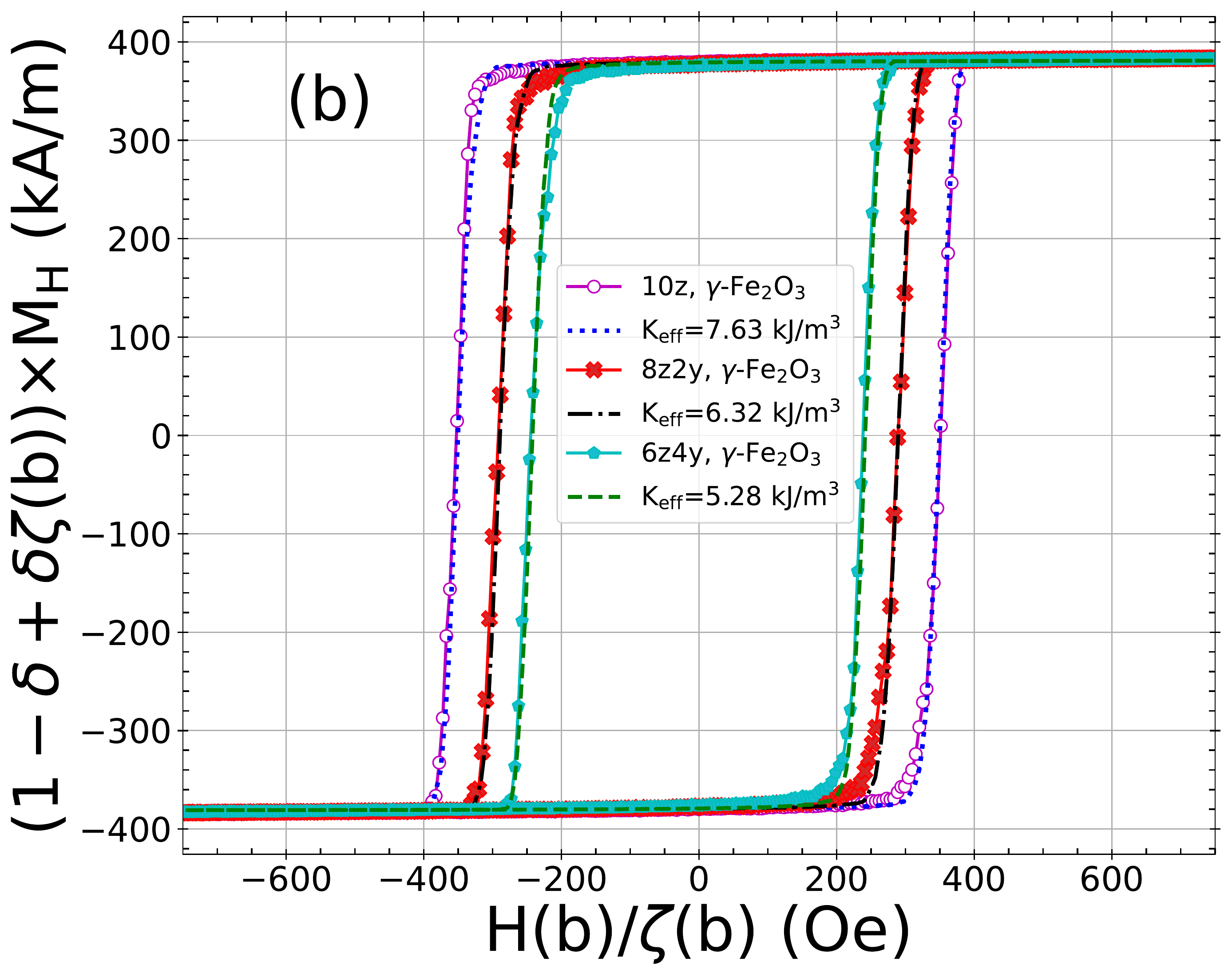}
        \caption{Impact of changing the field direction on loops for NPs composed of rods and equivalent macrospin (MS) particles. 
        a)  MS particle with $K_{\mathrm{eff}}$=4.8 kJ/m$^3$ and $M_s \simeq$382 kA/m has the same $M_r$ and $H_c$ as a $6z4y$ magnetite NP ($K_0=0$) under rotationally averaged field (black curve, circles, labelled SW in legend), 
        whereas it exhibits different MH hysteresis loops when field is applied along the $x$, $y$ or $z$ axes. For the filed applied along the $z$ axis, an MS with $K_{\mathrm{eff}}$=3.64 kJ/m$^3$ yields a similar hysteresis loop to the NP. b) Field along $z$, the effective anisotropy of $6z4y$ maghemite NP decreases (relative to a rotationally averaged field) from 5.7 to 5.28~kJ/m$^3$, for $8z2y$ NP it decreases from 8.78 to 6.32~kJ/m$^3$and for $10z$ maghemite NP from 10.79 to 7.63~kJ/m$^3$.   }\label{fig:NP_Hz_Keff}
\end{figure}

As with the case of individual rods, it is expected that a single anisotropy axis is not completely sufficient to model the magnetic response.  Fig.~\ref{fig:NP_Hz_Keff}a shows the response of the $6z4y$ magnetite nanoparticle model to both rotationally averaged fields and for fields along $x$, $y$ and $z$ directions, along with corresponding responses of the SW macrospin model that best matches the rotationally averaged response of the nanoparticle ($K_{\rm eff}=4.8$~kJ/m$^3$).  The nanoparticle loops for the $x$ and $y$ directions are  non-linear at moderate field magnitudes and have non-zero loop areas, while the macrospin model shows linear response until saturation and zero loop area.  Also shown is the loop for the macrospin model with a reduce effective anisotropy ($K_{\rm eff}=3.64$~kJ/m$^3$) that best matches the nanoparticle's repsonse to a field in the $z$ direction.  Fig.~\ref{fig:NP_Hz_Keff}b shows that lower values of $K_{\rm eff}$ are need to reproduce the response of maghemite nanoparticles to fields along $z$. The equivalent effective anisotropy under $H_z$  decreases to 5.28, 6.32, and 7.63 kJ/m$^3$ for $6z4y$, $8z2y$, $10z$ maghemite nanoparticles, respectively.  Effective parameters are summarized in Tab.~\ref{table:Keff}.  The up to approximately 35\% difference in $K_{\rm eff}$ values comparing rotationally average and $z$ responses can either be regarded as a model error when using the macropsin model for future purposes, or one may preferentially choose one scenario over the other depending on context.  For example, in a medium in which the nanoparticles are free to rotate and therefore can align anisotropy axes along the field, the lower $K_{\rm eff}$ values obtained from the $z$ response should be used, while for randomly oriented particles unable to rotate, the rotationally averaged may be more relevant.


\begin{table}
\caption{The effective anisotropy and saturation magnetization of macrospins equivalent to simulated nanorods and nanoparticles. The bulk saturation magnetization is $M_s=480$ kA/m.}
\centering
\begin{tabular}{|m{1.5 cm} |m{1.5cm} | m{1.4 cm}| m{1.7 cm} |m{1.2 cm}|}
\hline
material \newline & object \newline & $K_{\mathrm{eff}}$ (kJ/m$^3$) & $H$ \newline & $M_s^{\rm eff}$ (kA/m) \\
\hline
 Fe$_3$O$_4$ & nanorod & 15.73 & rot. avg. & 350 \\
\hline
 Fe$_3$O$_4$ & nanorod & 15.0 & $ ||z$ & 384 \\
\hline
$\gamma$-Fe$_2$O$_3$ & nanorod & 19.4 & rot. avg. & 350 \\
\hline
$\gamma$-Fe$_2$O$_3$ & nanorod & 18.7 & $||z$ & 384 \\
\hline
Fe$_3$O$_4$ & $6z4y$ NP & 4.80 & rot. avg. & 382 \\
\hline
Fe$_3$O$_4$ & $6z4y$ NP & 3.64 & $||z$ & 382 \\
\hline
$\gamma$-Fe$_2$O$_3$ & $6z4y$ NP & 5.70 & rot. avg. & 382 \\
\hline
$\gamma$-Fe$_2$O$_3$ & $6z4y$ NP & 5.28 & $||z$ & 382 \\
\hline
$\gamma$-Fe$_2$O$_3$  &$8z2y$ NP & 8.78 & rot. avg. & 382 \\
\hline
$\gamma$-Fe$_2$O$_3$  &$8z2y$ NP & 6.32 & $||z$ & 382 \\
\hline
$\gamma$-Fe$_2$O$_3$  &$10z$ NP & 10.79 & rot. avg. & 382 \\
\hline
$\gamma$-Fe$_2$O$_3$  &$10z$ NP & 7.63 & $||z$ & 382 \\
\hline
\end{tabular}\\
\label{table:Keff}
\end{table}


\section{\label{sec:TwoNPs}Interacting Nanoparticles }

\begin{figure*}
    \centering
    \includegraphics[width=0.51\textwidth]{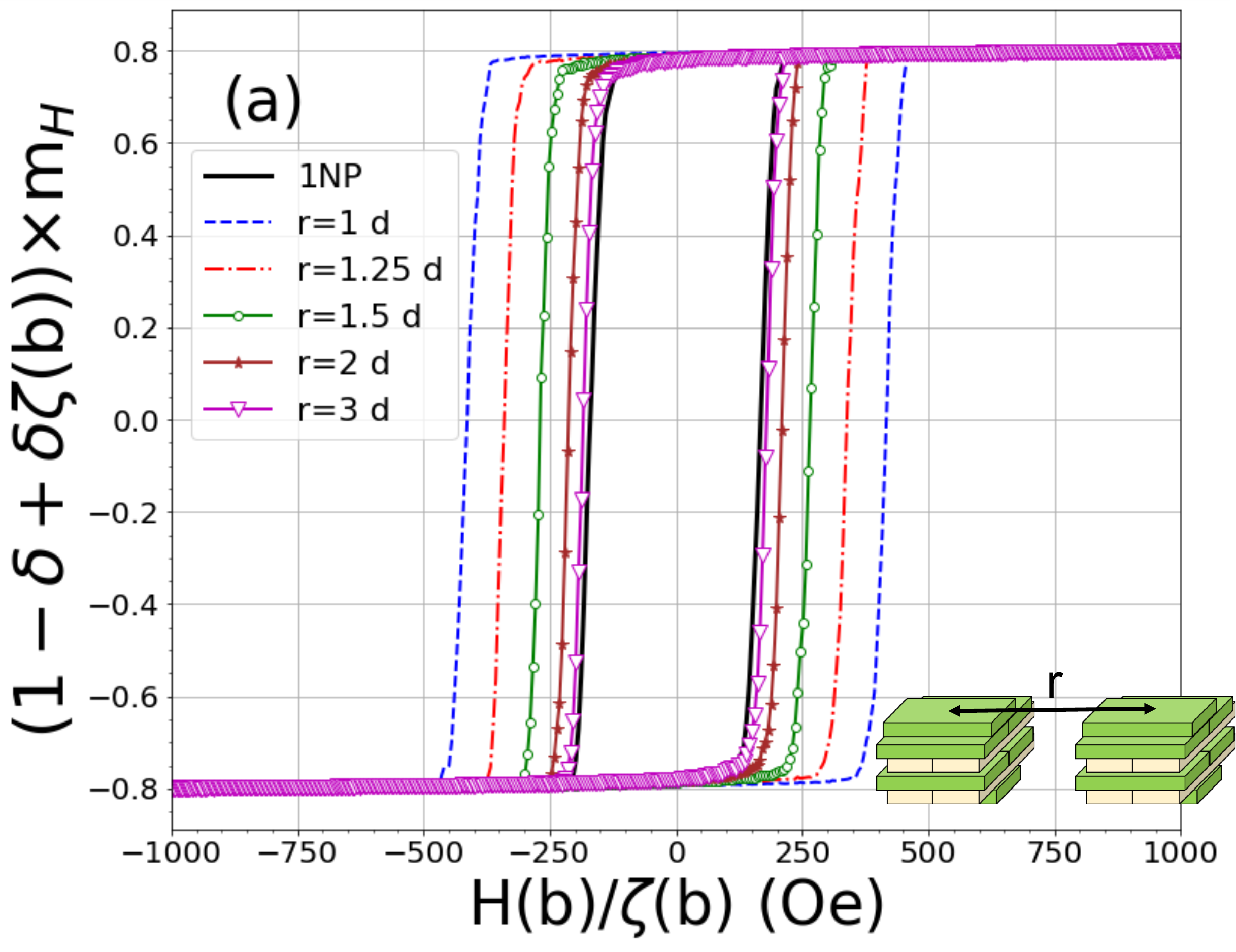}
    \includegraphics[width=0.48\textwidth]{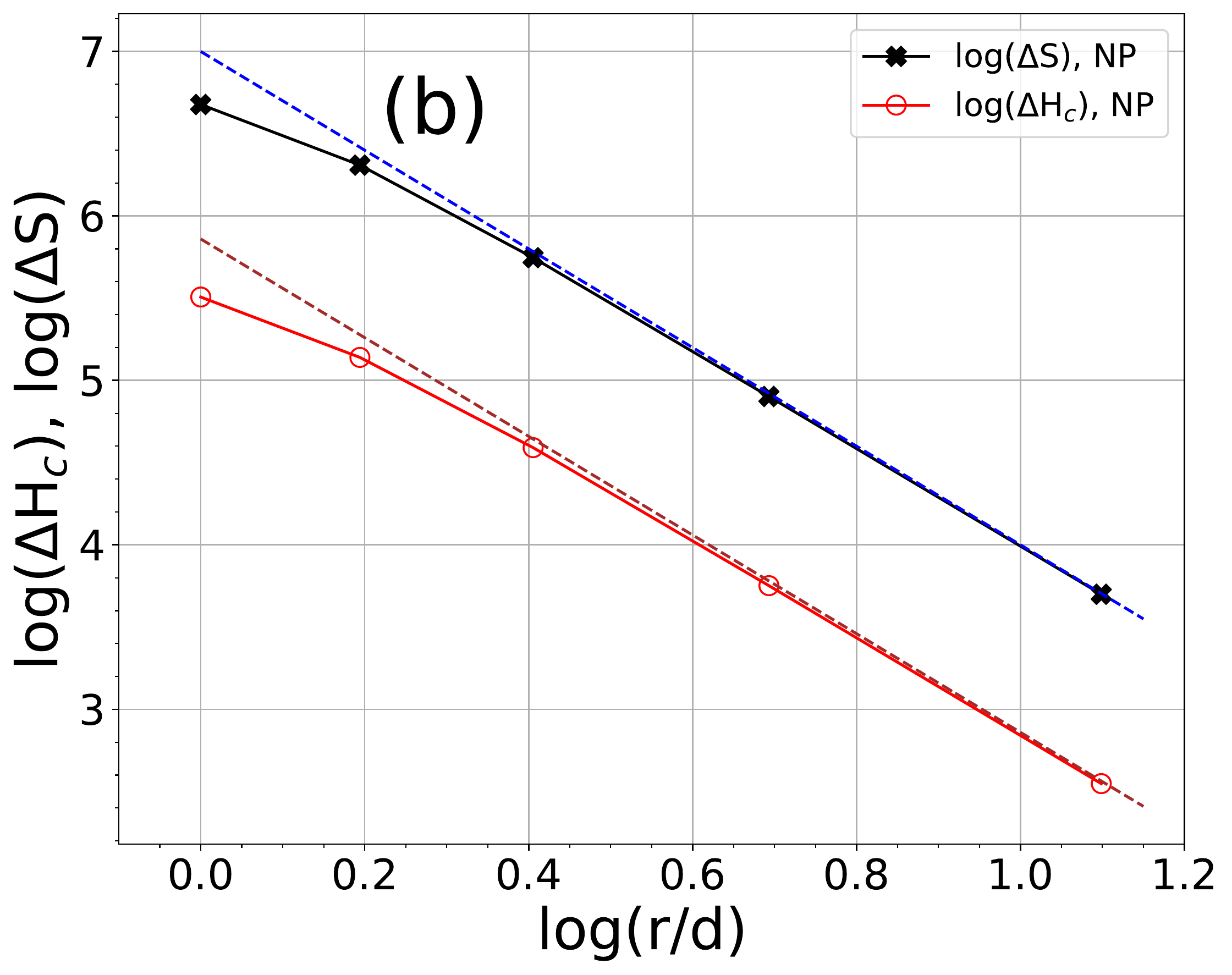}
    \includegraphics[width=0.48\textwidth]{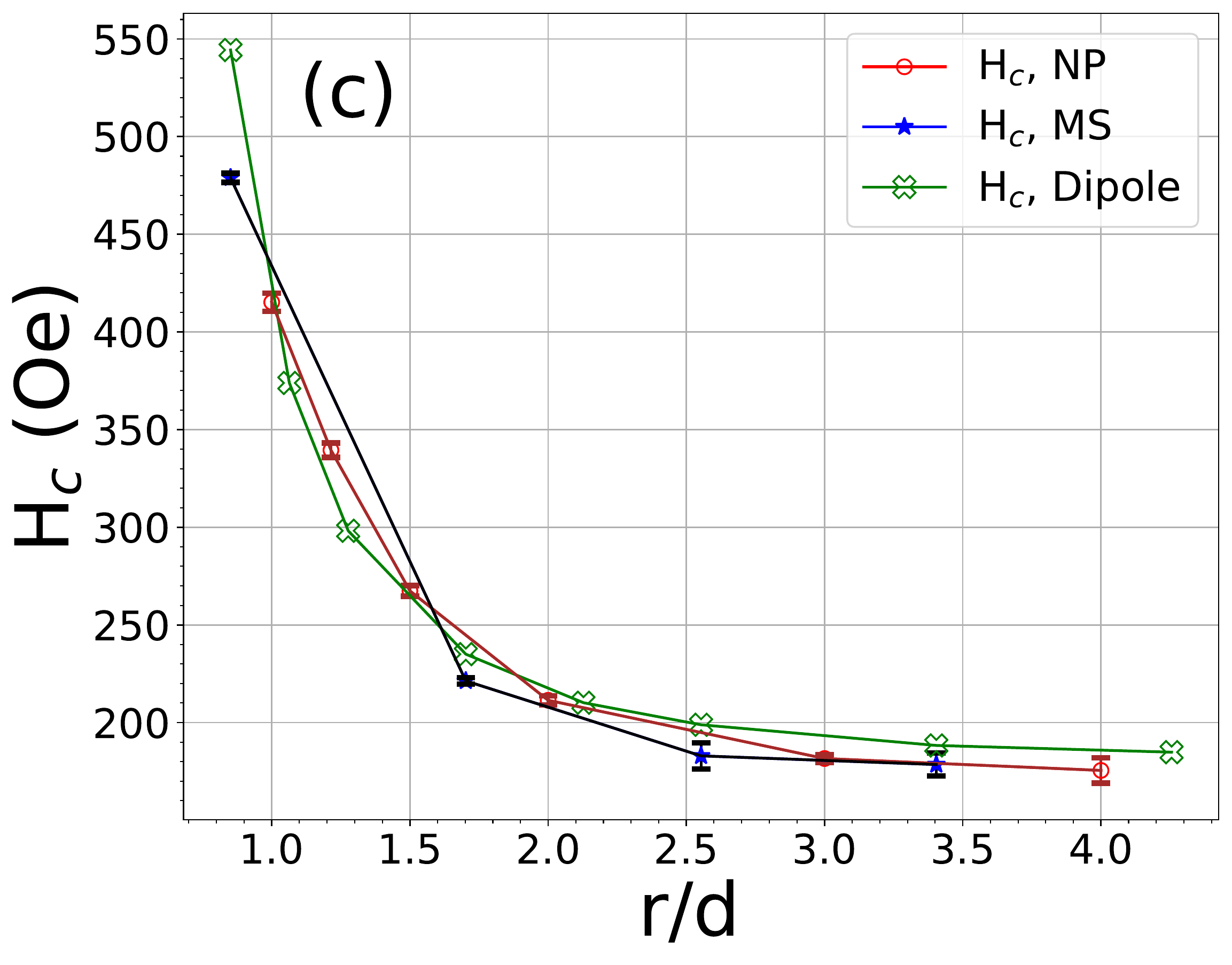}
    \caption{a) Hysteresis loops for a system of 2 magnetite $6z4y$ NPs as a function of centre-to-centre distance $r$. $d$ is the NP diameter. 
    b) The quantities $\Delta H_c$ and $\Delta S$ (see main text for definitions) as functions of $r$ approach dipolar scaling near $r/d=1.5$ ($\ln1.5 \approx 0.405$).  Dashed lines are $1/r^{-3}$ power laws.
    (c)  $H_c$ as a function of $r$ for the 2-NP loops from panel (a), along with $H_c$ obtained from macrospin approximations to the NPs, realized through uniformly magnitized cubes (MS) and dipolar spheres (Dipole).  Error bars for the dipole curve are comparable to symbol size.
    }
    \label{fig:SLP_r2NP}
\end{figure*}


As a prelude to later explorations of the collective heating behavior of NP chains, as in Ref.~\cite{bakuzis2013chain}, we simulate two magnetite $6z4y$ NPs ($K_0=0$) and study how their hysteresis loop changes as the nanoparticle center-to-center distance $r$ varies from one to three NP diameters ($d=47.0$~nm).  For these simulations, the centres and anisotropy axes of both NPs lie on the $z$ axis, the external field is also along $z$, and we use $b=4$ for coarse-graining (including magnetostatic scaling). This arrangement mimics chain formation when NPs are free to move and rotate.  As shown in Fig.~\ref{fig:SLP_r2NP}a, the hysteresis loop area is larger in the case of two interacting chained NPs compared to isolated NPs.  This is in agreement with reported results~\cite{cabrera2018dynamical, serantes2014multiplying, mehdaoui2013increase}.  We note that the normalization of the loop is such that the total heat released would require multiplication by the number of particles in the system.  As $r$ increases, the effect of magnetostatic interactions between NPs is reduced and their loop area shrinks\cite{torche2020thermodynamics}.  By $r\approx 3d$, the loop is approximately the same as for noninteracting NPs.

To quantify the $r$ dependence of the loop area and $H_c$, we plot the difference in areas $\Delta S$ between loops for the 2-NP systems and individual NPs ($\Delta S=$ Area(2NPs) - Area(1NP)), as well as the difference in the coercivities $\Delta H_c$, as functions of $r$ in Fig.~\ref{fig:SLP_r2NP}b.  
As may be expected, for $r> 1.5 d$, $\Delta S$ and $\Delta H_c$ decrease with a $1/r^3$ dependence, just as the energy between two dipoles does. This motivates using the dipole approximation to calculate the heating efficiency of NPs when they are further apart than $1.5 d$. 

To this end, we carry out two additional sets of simulations.  First, we use the effective macrospin parameters for the $6z4y$ magnetite NP ($K_{\mathrm{eff}}=3.64$ kJ/m$^3$, $M_s$=381.6 kA/m) and simulate two magnetized cubes with the same volume as the NP, placing their centres and anisotropy axes on the $z$ axis, and calculating loops as we vary $r$.  For these simulations, we include magnetostatics interactions, both between the cubes and within each cube.  Allowing for self-demagnetization is technically inconsistent with our approach, but the self-demagnetization leads to cubic anisotropy that has little effect on hysteresis loops.  Similarly, we simulate with Vinamax software~\cite{leliaert2015vinamax} two spheres with the same effective parameters as the cubes and dipole moment $v M_s^{\rm eff}$, thus neglecting self-demagnetization (as is consistent with the effective parameters) and treating interaction between spheres in the dipolar approximation.  We report $H_c$ for the two macospin models and the $6z4y$ NPs in Fig.~\ref{fig:SLP_r2NP}c, with cubes labelled {\it  MS } and spheres labelled {\it Dipole}.  The agreement between all three sets of data is satisfactory for $r\ge 1.5 d$.



\section{Conclusions}\label{sec:sonclusion}


The present work represents the first comprehensive study of coarse-graining for use in micromagnetic simulations. We extend an RG-based coarse-graining scheme, previously developed and explored in I, to include magnetostatic interactions in micromagnetic simulations, and apply it to dynamic hysteresis loops at $T=310$~K of magnetite (no magnetocrystalline uniaxial anisotropy) and maghemite nanorods, as well as collections of stacked nanorods that model NPs of varying internal orientational order.

For individual nanorods, the coarse-graining procedure reproduces loops even up to the representation of the nanorod as a block with a single magnetization.  For collections of rods, the interplay between inter-rod exchange and magnetostatic interactions can lead to complex magnetization dynamics, and we limit our level of coarse graining to $b=4$ (cell length four times larger than the unit cell of magnetite) when simulating 10-nanorod model NPs.

For both individual nanorods and NPs, we find the effective uniaxial anisotropy and saturation magentization parameters for SW macrospin models that yield equivalent loops.
For nanorods, the effective anisotropy is approximately 15-16~kJ/m$^3$ for magnetite, and approximately 19~kJ/m$^3$ for maghemite.  The effective saturation magnetization is 73\% to 80\% of the bulk value, depending on whether orientation with respect to the external field is assumed to be rotationally averaged or parallel. 
For our 47~nm-diameter NPs, the effective anisotropy falls in the range of 4~kJ/m$^3$ for our most orientationally disordered ($6z4y$) magnetite NP to 11~kJ/m$^3$ for our most ordered ($10z$) maghemite NP.  The effective saturation magnetization is approximately 80\% of the bulk value.  For this modelling, we assume an inter-rod exchange strength of half the bulk value.

For simulations of two NPs, we find that loop area, or rather the difference in loop areas between interacting and noninteracting NPs, scales with distance in a dipole-like manner for centre-to-centre distances at and beyond 1.5 times the particle diameter.  For this distance and beyond, we find good agreement between the two-NP results and those for two macrospin equivalents interacting via dipolar interactions.

Thus, our methodology starts with micromagnetic parameters at the unit cell level, and, through coarse-graining, allows us to carry out micromagnetic simulations of nano-sized particles with properly scaled parameters and magnetostatic interactions.  Further, we find equivalent macrospin models with effective anisotropies and saturation magnetizations that can be used in micromagnetic or molecular dynamics simulations involving a large number of NPs.  This knowledge allows for the simulation of larger systems with more detail than is normally assumed within macrospin models and should be extendable to non-hyperthermia applications.

We also find (Appendix A) that using a larger cell size allows the use of a larger step size in integrating the equations of motion.  Over the range of cell sizes studied, we approximately find that if cell volume is increased, the step size may also be increased by the same factor.

\section*{Acknowledgment}
We thank Michael J. Donahue for discussions and his expert guidance on how to achieve the scaling of magnetostatics with OOMMF. We also thank Mikko Karttunen for useful discussions, and both him and Styliani Consta for hosting our stay (RB and ISV) at Western University.  We acknowledge the financial support from the Natural Sciences and Engineering Research Council (Canada).  Computational resources were provided by ACENET and Compute Canada.

\appendix
\section{Time step dependence on simulation cell size}\label{app:time_stp}

\begin{figure*}\centering
\includegraphics[width = 0.45 \textwidth]{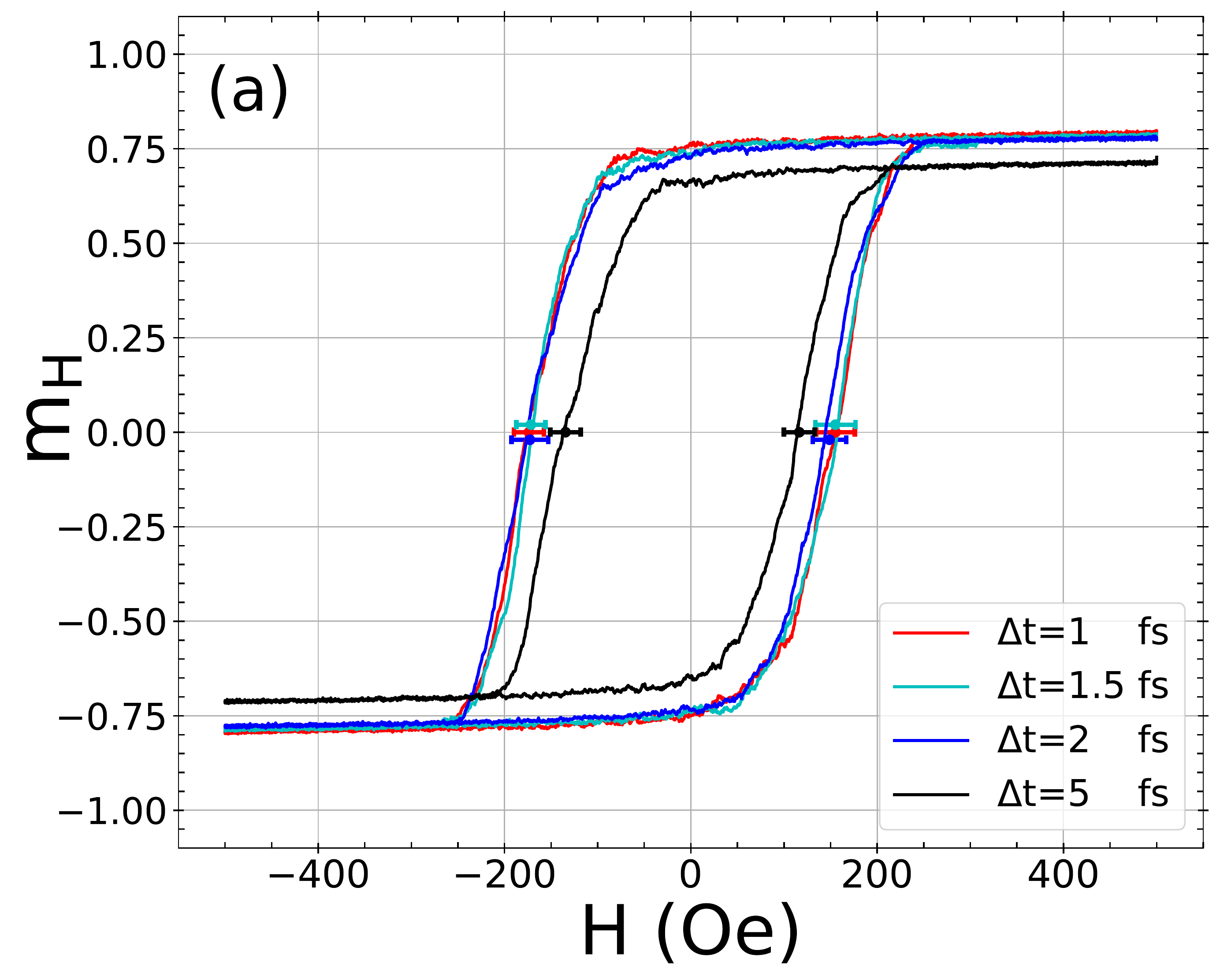}
\includegraphics[width = 0.45 \textwidth]{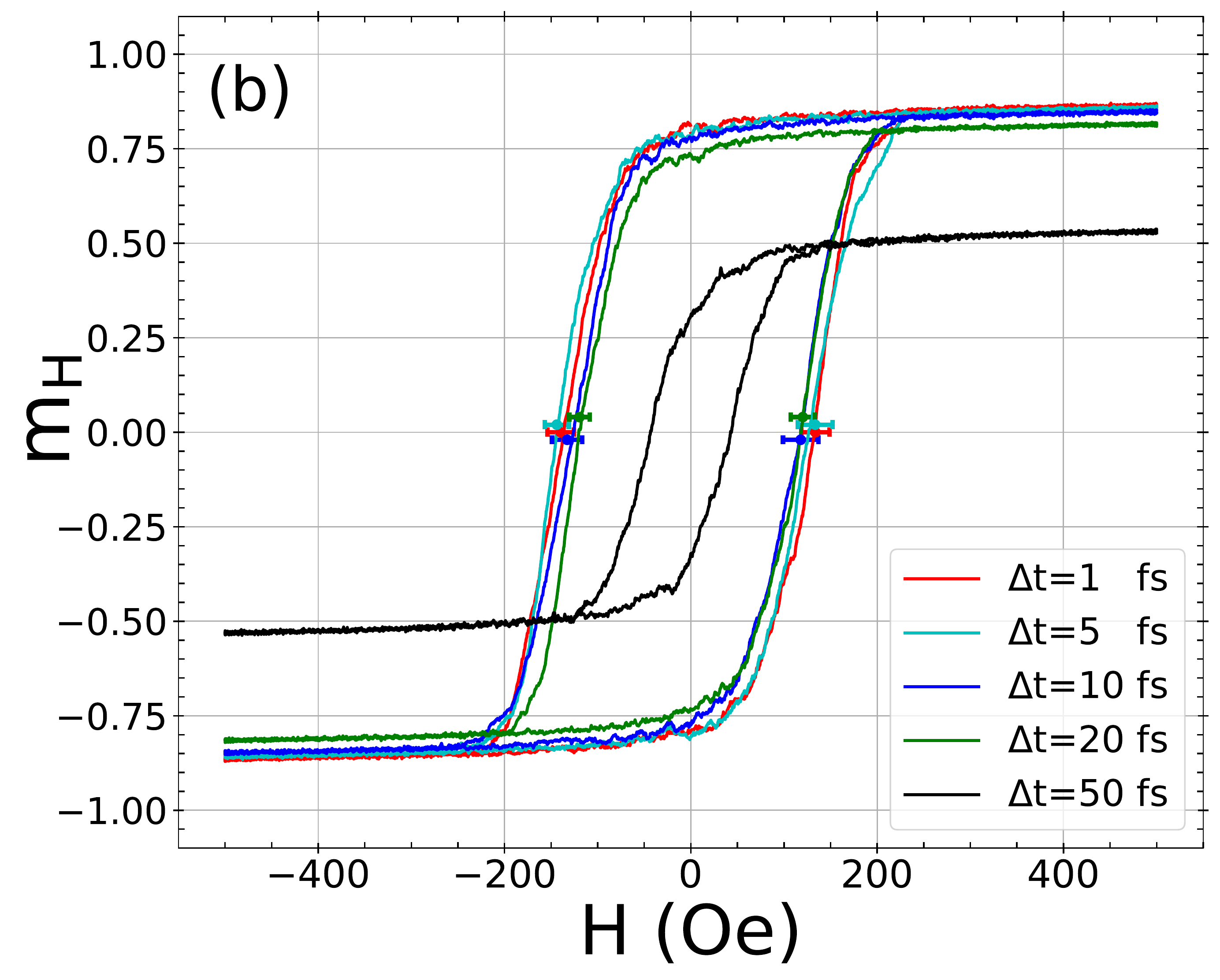}
\includegraphics[width = 0.45 \textwidth]{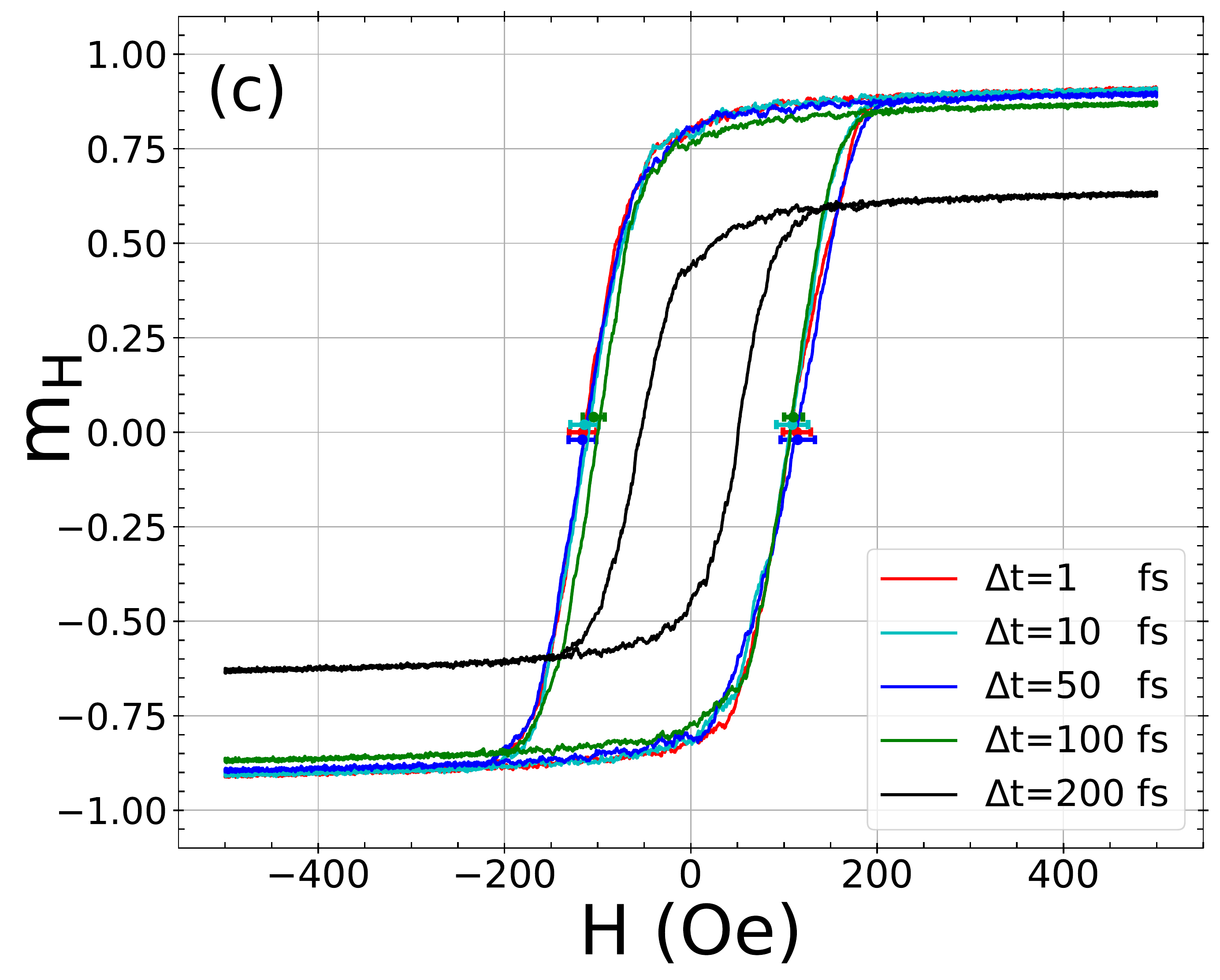}
\includegraphics[width = 0.45 \textwidth]{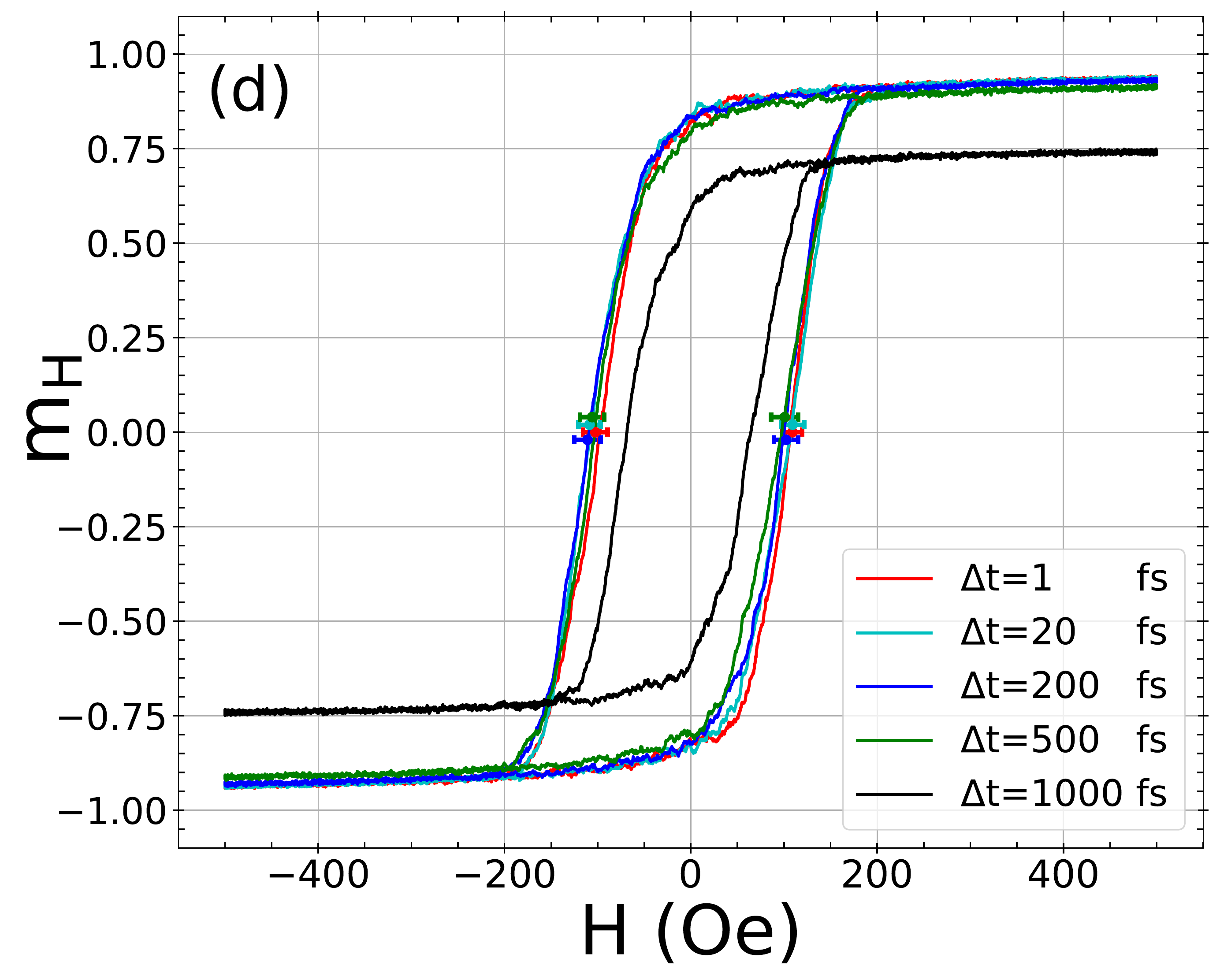} 
\caption{Dependence of MH loops on $\Delta t$ for nanorods composed of cells of side length $b a_0$ for (a) $b=1$, (b) $b=2$, (c) $b=4$ and (d) $b=8$. The simulations are carried out at SR=2.5~Oe/ns, and $T=310$~K, with $\alpha=0.1$.  Here, we neglect magnetostatic interactions.
}
\label{fig:step_cell}
\end{figure*}



For a system of interacting micromagnetic cells with crystalline anisotropy under an external field, the Hamiltonian is,
\begin{equation}
\begin{split}
    &\mathcal{H}=\mathcal{H}_{\mathrm{exchange}} + \mathcal{H}_{\mathrm{magnetostatics}} + \mathcal{H}_{\mathrm{anisotropy}} + \mathcal{H}_{\mathrm{Zeeman}}\\
    & \mathcal{H}_{\mathrm{exchange}}= -\frac{a }{2}\displaystyle\sum_{i}\displaystyle\sum_{j \in{\rm NN}}A_{ij} (\mathbf{m}_i\cdot\mathbf{m}_j)\\
    & \mathcal{H}_{\mathrm{magnetostatics}}=- \frac{\mu_0v}{2} \displaystyle\sum_{i,j}(\mathbf{M}_i\cdot\mathbf{N}\cdot\mathbf{M}_j) \\
    & \mathcal{H}_{\mathrm{anisotropy}}=-K v \displaystyle\sum_i(\mathbf{m}_i\cdot\mathbf{u})^2\\
    & \mathcal{H}_{\mathrm{Zeeman}} = -\mu_0M_sv\displaystyle\sum_i (\mathbf{m}_i\cdot \mathbf{H})
\end{split}
\end{equation}
where $a$ is the length of a cubic cell and $A_{ij}$ is the exchange stiffness constant. We note that  the factor of 1/2 in the exchange Hamiltonian may or may not appear in the literature, reflecting whether or not interactions are effectively double counted, resulting in the apparent values of $A_{ij}$ differing by a factor of 2.  For example, for magnetite we use a value of $A_0=0.98\times 10^{-11}$J/m, and to achieve this we give as input the parameter $A_{\rm OOMMF}=0.49\times10^{-11}$~J/m to OOMMF.
$\mathbf{M}_k$ is a cell's magnetization  with magnitude $M_s$ and direction given by unit vector $\mathbf{m}_k$ ($k=i,j$), $\mathbf{N}$ is the demagnetization tensor, representing the geometry of the system, $\mu_0$ is the permeability of the free space and $v$ is the cell volume. Uniaxial anisotropy is characterized by energy density $K$ and unit vector $\mathbf{u}$, and the externally applied field is $\mathbf{H}$.  

Brown~\cite{brown1963thermal} modelled thermal effects with a random effective field (white noise) with spatial components drawn from a normal distribution with variance~\cite{leliaert2015vinamax},
\begin{equation}\label{eq:thermal}
    \sigma^2 = \frac{2\alpha k_B T}{\gamma \mu_0 M_s V\Delta t},
\end{equation} 
where $V$ is the switching volume, i.e. the volume of a micromagnetic cell, $T$ is the absolute temperature, $k_B$ is Boltzmann's constant, and $\Delta t$ is the time step of the simulations. Eq.~\ref{eq:thermal} implies that a larger $\Delta t$ can be chosen for larger simulation cells.  Therefore, when coarse-graining, not only are simulations faster on account of employing fewer cells, but also on account of being able to use a larger $\Delta t$.  In Fig.~\ref{fig:step_cell} we plot hysteresis loops for nanorods composed of cells with different volumes, given by $V=(b a_0)^3$, for different values of $\Delta t$. 
For these simulations, we use the RG-scaled exchange and anisotropy constants $A(b)$ and $K(b)$ as given by Eqs.~\ref{eq:RNG_scalingA} and~\ref{eq:RNG_scalingK}; we also neglect magnetostatic interactions for simplicity. Overlapping curves indicate that results are independent of step size, and therefore indicate when $\Delta t$ is ``small enough''.
For $b=1$, a small $\Delta t$ of approximately 1 to 1.5~fs is required, and the optimal $\Delta t$ increases to approximately 5~fs for $b=2$, 50~fs for $b=4$ and, remarkably, 200~fs for $b=8$. 
Values of $\Delta t$ larger than the optimum yield significantly smaller loop areas.
This increase of time step with cell volume is consistent with previous results in the literature~\cite{lopez2012micromagnetic,kapoor2006effect}.

OOMMF uses an Eulerian solver for simulations at finite $T$, and so the contribution to the changes in magnetization from the thermal field in a single step of the algorithm is proportional to $\sqrt{\Delta t/V}$, which implies that for $\Delta t \propto b^3$ the magnitude of these changes should remain constant. This proportionality provides a simple way of understanding the increase in optimal $\Delta t$ that we observe.  
It should be cautioned, however, that care must always be taken to check that a sufficiently small $\Delta t$ is used.

\bibliographystyle{apsrev4-2}

\bibliography{ref-num.bib}

\end{document}